\documentclass[reprint,amsmath,amssymb,aps,pre,superscriptaddress]{revtex4-1}

\usepackage{graphicx}
\usepackage{dcolumn}
\usepackage{soul}
\usepackage{bm}
\usepackage{mathptmx}
\usepackage{textcomp} 
\usepackage{xcolor}
\newcommand{\avg}[1]{\langle #1 \rangle}

\begin{document}

\preprint{APS/123-QED}

\title{Yielding, rigidity, and tensile stress in sheared columns of hexapod granules}
\author{Yuchen Zhao}
\email{yz172@phy.duke.edu}
\affiliation{Department of Physics, Duke University, Durham, North Carolina, USA}
\author{Jonathan Bar\'{e}s}
\affiliation{Laboratoire de M\'{e}canique et G\'{e}nie Civil, UMR 5508 CNRS-University Montpellier, 34095 Montpellier, France}
\author{Joshua E.~S.~Socolar}
\affiliation{Department of Physics, Duke University, Durham, North Carolina, USA}

\date{\today}

\begin{abstract}
Granular packings of non-convex or elongated particles can form free-standing structures like walls or arches.  For some particle shapes,  such as staples, the rigidity arises from interlocking of pairs of particles, but the origins of rigidity for non-interlocking particles remains unclear.  We report on experiments and numerical simulations of sheared columns of ``hexapods,'' particles consisting of three mutually orthogonal sphero-cylinders whose centers coincide.  We vary the length-to-diameter aspect ratio, $\alpha$, of the sphero-cylinders and subject the packings to quasistatic direct shear.  For small $\alpha$, we observe a finite yield stress.  For large $\alpha$, however, the column becomes rigid when sheared, supporting stresses that increase sharply with increasing strain.  Analysis of x-ray micro-computed tomography (micro-CT) data collected during the shear reveals that the stiffening is associated with a tilted, oblate cluster of hexapods near the nominal shear plane in which particle deformation and average contact number both increase.  Simulation results show that the particles are collectively under tension along one direction even though they do not interlock pairwise.  These tensions comes from contact forces carrying large torques, and they are perpendicular to the compressive stresses in the packing.  They counteract the tendency to dilate, thus stabilizing the particle cluster.

\end{abstract}

\maketitle

\section{\label{sec:intro}INTRODUCTION}
An important challenge in the science of granular materials is to understand the connection between the shapes of individual grains and the macroscopic response of the aggregate~\cite{jaeger2015_sm}. Recent studies have shown that nontrivial desired macroscopic material properties can be obtained by tuning the grain shape~\cite{jaeger2015_sm,Keller2016_gm,Borzsonyi2013_sm,weiner2020_jap,damme2018_concrete}. For non-cohesive particles, spherical or nearly spherical shapes form packings that deform plastically under shear \cite{azema2007_pre,azema2010_pre,azema2013_pre,athanassiadis2014_sm,murphy2019_prx}. However, packings of highly elongated and/or strongly non-convex particles show stiffening behavior under shear \cite{philipse1996_lang,desmond06_pre,Blouwolff2006_epl,Franklin2014_epl,zhao2017_pg,bares2017_pg,guo2020_AIChE}. A dramatic illustration of this effect is the formation of free standing walls and columns consisting of slender rods, staples, granular chains, or star-shaped particles~\cite{trepanier2010_pre,gravish2012_prl,brown2012_prl,Zhao2016_gm,Fauconneau2016_gm}.
By analogy to similar properties of wet sand, in which water bridges provide cohesive forces between grains, dry granular materials that support such structures in the absence attractive interaction between grains are said to exhibit ``geometric cohesion''~\cite{franklin2012_pt}.
Columns of dry granular materials exhibiting geometric cohesion can have a large yield stress under uniaxial compression~\cite{brown2012_prl,Murphy2016_gm,dumont2018_prl}.

Fundamental questions remain open regarding the microscopic sources of geometric cohesion. Previous research has focused on the effect of entanglement in packings of highly non-convex particles \cite{gravish2012_prl,brown2012_prl,dumont2018_prl}. Staples, for example, can act like hooks to form interlocking chains that resist tension \cite{gravish2012_prl, Franklin2014_epl}. However, it is not clear how particles manage to form a cohesive or stiffening packing when the particle shape does not allow a single pair of grains to support tensile stress. In addition, the implications of geometric cohesion for elastic and rheological properties are not well understood. What configurations of noncohesive particles provide the tensile stresses required to avoid dilation and thereby resist large applied stresses? And in cases where the material has a finite yield stress, does geometric cohesion give rise to yield stress \textit{vs.} pressure curves similar to those produced by wet granular materials?  In addition to their intrinsic interest, these questions are highly relevant for civil and material engineering applications \cite{dierichs2015_ad,Keller2016_gm,damme2018_concrete}.

This paper reports on direct shear experiments and numerical simulations with aggregates of ``hexapods'' which are particles shaped as shown in Fig.~\ref{fig:setup}. Each particle consists of six cylindrical arms of equal length emanating from a center along three mutually perpendicular directions. We define $\alpha$ to be the ratio of the length of the particle diameter ($2$ arm lengths) to the diameter of a cylindrical arm. For the packings comprised of particles with $\alpha$ near unity, we observe plastic yielding of the granular material at finite yield stresses. For large $\alpha$, the material stiffens and does not yield before individual particles break. We use x-ray micro-CT to measure the bending of particle arms and identify a rigid cluster of particles that is responsible for supporting the applied stress. We also perform numerical simulations on hexapods for two values of $\alpha$, finding good agreement with experiments, and use the simulations to identify the source of the tensile stresses that counterbalance the tendency toward dilation and prevent plastic yielding.

The rest of this paper is organized as follows. In Sec.~\ref{sec:expt}, we describe the direct shear experimental setup and the x-ray micro-CT data acquisition system. In Sec.~\ref{sec:exptres}, we present the experimental stress-strain curves and analyze the associated packing structures. In Sec.~\ref{sec:simu}, we present numerical simulations that show qualitative behavior similar to that observed in our experiments and analyze the simulated contact forces to identify the structure that leads to tensile stresses. Section~\ref{sec:conclusion} contains a discussion and concluding remarks.

\section{\label{sec:expt}EXPERIMENTS}

\subsection{\label{sec:exptsetup}Experimental setup and procedures}
\begin{figure}
 \centering
  \includegraphics[width=3.375 in,keepaspectratio=true]{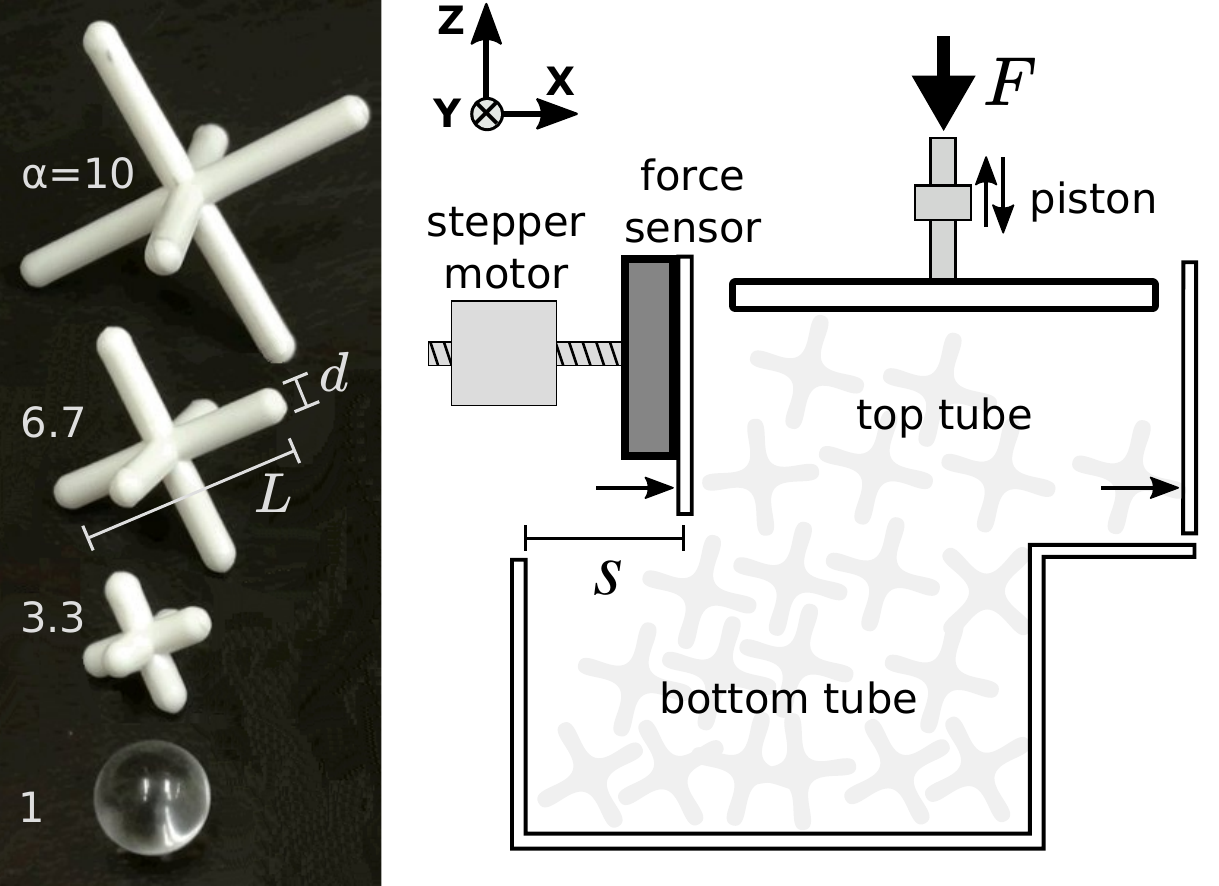}
 \caption{(Left) Spherical and hexapod particles used in the experiments. (Right) A schematic of the direct shear cell.}
 \label{fig:setup}
\end{figure}

Shown in Fig.~\ref{fig:setup}, the experimental apparatus is a direct shear cell of a type commonly used in granular and soil material testing \cite{Wood1990_book1}. It consists of two stacked acrylic cylindrical tubes of diameter $D=96\,$mm. The bottom tube is fixed to a base and has a height of $85\,$mm. The top tube sits on a linear guide, which is supported by the same base and permits horizontal motion in one direction, which we define to be the $x$ direction. The tubes are separated by a small vertical gap ($\approx1\,$mm) compared to the particle size. A stepper motor drives the top tube in the shear direction at $0.1\,$mm/s. A force sensor (strain-gauge load cell) connects the top tube and the motor, and measures shear force with $0.1\,$N accuracy at a frequency of $1\,$Hz. A piston, which is held by another linear guide attached to the top tube, can be used to apply constant normal force on the top of the packing throughout the shearing process. Another support extends horizontally in the shear direction to prevent particles from falling out of the top tube at large shear strains. We cover this support with a low friction Teflon sheet to reduce frictional drag on the particles during the shear.

Our experiments are conducted using plastic (polypropylene) hexapods that consist of three mutually orthogonal cylinders with spherical caps, whose centers coincide (see the left panel of Fig.~\ref{fig:setup}). The diameter $d$ and lengths $L$ of the cylinders are $3$, $10$, $20$ and $30\,$mm, giving them a length-to-diameter aspect ratio $\alpha=3.3,\,6.7,\,10$. The material has Young's modulus on the order of $1\,$GPa and a static friction coefficient $0.36\pm0.05$. We also use acrylic spheres with diameter $L=9.5\,$mm as a benchmark, and it has $\alpha=1$.

For each experiment, a monodisperse packing is prepared by randomly pouring particles into the initially aligned tubes and then leveling the top surface of the packing. The initial height of the packing ranges from $140$ to $150\,$mm for different $\alpha$, which fills the entire bottom tube and part of the top tube. A nominal initial packing fraction, defined as the particle volume divided by the volume of the cylindrical region they occupied, is $0.59\pm0.02$ ($\alpha=1$), $0.40\pm0.04$ ($\alpha=3.3$), $0.23\pm0.01$ ($\alpha=6.7$) and $0.14\pm0.01$ ($\alpha=10$). 

In a given run of the experiment, a constant normal force $F$ is applied through the piston, and the stepper motor drives the top tube in the $x$ direction at constant speed, which generates a shear force, continuously measured by the force sensor. The evolution of this horizontal force is recorded for several values of $F$, taking several runs at each value. Gravity contributes to the normal stress on the plane, and we define a nominal normal stress $P=(F+G)/A$, where $G$ is the total gravitational force on the particles above the shear plane $z=0$, and $A=\pi D^2/4$ is the tube's cross-sectional area. We note that in addition to $P$, there is a component of normal stress associated with frictional forces applied by the tube walls to the particles. We define the shear stress $\tau$ to be the shear force divided by the area of intersection of the top and bottom tubes at $z=0$. The shear strain $\gamma$ is defined as $s/L$, where $s$ is the top tube displacement measured by counting the steps taken by the motor (see the right panel of Fig.~\ref{sec:exptsetup}). We stop the shear when $s=30\,$mm or the force on the force sensor exceeds $20\,$N.

\subsection{\label{sec:ctscanpostprocess}x-ray micro-CT data acquisition and post-processing}
We use an x-ray micro-CT scanner (Nikon XT-H225) to observe packing structures of $\alpha=10$ particle packing under shear. Three repeated and independent runs are done to check the consistency of our observations. The packings are prepared in the same way as in Sec.~\ref{sec:exptsetup}, without the piston to apply addition normal stress. Each run is paused at different $\gamma$ to take an x-ray scan, during which the tubes are removed from the force sensor and motor with the top tube clamped to the linear guide to resist force from the packing. During an x-ray scan, the sample is very slowly rotated along a vertical axis to collect projection images, with a x-ray source of about $190$~kV and $180$~$\mu$A. These projections are then post-processed using Nikon's Feldkamp cone based CT algorithm \cite{Feldkamp1984_JOSA} to get $16$-bit 3D density image with size about $1500^3\,$px$^3$ and spatial resolution about $80$ \textmu m/px.

To extract packing structures from the 3D image, we use codes developed previously \cite{bares2017_pg}. Each 2D slice of the $16$-bit image is binarized using Otsu's method \cite{otsu1979_ieee}, producing a 3D density in which voxels occupied by material are set to $1$. We then calculate the Euclidean distance of each $1$ to its nearest $0$, and set to $0$ all of the $1$'s for which the distance is smaller than about $1/3$ of a particle arm diameter. The resulting connected regions of $1$'s correspond to individual particles, allowing for an estimate of the center of mass of each particle and the Euler angles specifying its orientation. We then use a template-matching technique on the original binarized image \cite{bares2017_pg,neudecker2013_prl} to refine these estimates. The template is taken to be an ideal hexapod with the appropriate dimensions. We check that all template overlap values are greater than $87\%$ and that no particle is missed. Overlaps of less than $95\%$ are attributable to the bending of hexapod arms in the physical sample. The position and orientation measurement accuracy are $1\,$px ($80$ \textmu m) and $0.3^{\circ}$ respectively.

\begin{figure}[tb]
 \centering
 \includegraphics[width=1.3 in,keepaspectratio=true]{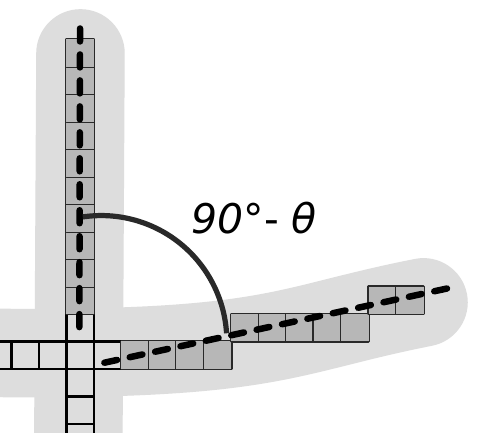}
 \caption{A schematic illustrates the bending of particle arm and measurement. The shaded light-gray area is the body of a particle. Boxes are one-pixel wide skeleton of the particle. The dashed lines are linear fits based on the coordinates of the skeleton of each arm. The bend angle $\theta$ is measured as $90^{\circ}$ subtracting the angle between the lines.}
 \label{fig:bendSch}
\end{figure}

Each particle from the eroded image is skeletonized to a width of $1$-voxel using an image thinning procedure \cite{Lee1994_skeleton} implemented in the Python package \texttt{scikit-image} \cite{scikit-image}. From the skeleton, we determine an angle between nominally orthogonal arms using the method illustrated in Fig.~\ref{fig:bendSch}. This measurement is a proxy characterizing the bending of particle arms. We fit each of the six arms (about $100$ voxels long) to a straight line using a least-squares method. The ``bend angle'' $\theta$ between nominally orthogonal arms is defined as $90^{\circ}$ minus the angle between the two straight lines. The error in $\theta$ is $0.4^{\circ}$ on average. Further, we define a quantity characterizing the total deformation of an individual particle $i$: 
\begin{equation}
\theta_i(\gamma)  = \bigg| \sum | \theta(\gamma)| -  \sum | \theta(0)|\bigg|\,
\end{equation} 
where the sums are taken over the $12$ pairs of orthogonal arms of a given particle, and $\gamma$ specifies the applied strain. The subtracted term accounts for any pre-existing distortions which are usually less than $0.3^{\circ}$.

Finally, we detect inter-particle contacts and particle-tube contacts using a previously developed technique \cite{bares2017_pg}. We first estimate contact locations from the skeletons. If the shortest distance between a pair of arm skeletons is smaller than $1.1d$, the midpoint of the shortest line segment connecting them is stored as a possible contact. We then zoom into a box of edge length $21\,$px ($\approx1.8\,$mm) centered at each possible contact in the original $16$-bit density image and binarize the image using the threshold taken from the binarization step discussed above. If this produces two disconnected domains, we conclude that there is no contact. Otherwise we take the original midpoint to be the location of a contact. We also vary the threshold within a reasonable range to determine the sensitivity of the contact detection. A maximum threshold is obtained by increasing the median threshold to a point where clearly identifiable contacts are missed, such as the contacts that support particles on top of the packing. A minimum threshold is taken to yield a range that is symmetric about the median. Varying the threshold can change the average contact number substantially. Nevertheless, the trend in average contact number with increasing strain is similar for all threshold choices, as will be shown below.

\section{Experimental Results} \label{sec:exptres}

We report here on measurements of the yield stress for packings with $\alpha = 1$ (spheres) and $3.3$, and on the nature of the geometric form of the network of particles that support strong macroscopic stresses for $\alpha = 6.7$ or $10.0$, where the packing stiffens rather than yielding to applied stresses that would cause particles to break.

Plastic yield of slowly sheared granular materials occurs above a threshold shear stress $\tau$, which generally depends on the normal stress $P$ perpendicular to the shear plane~\cite{andreotti2013_book}. In many cases, $\tau$ depends linearly on $P$: $\tau=\mu P+c$, where the constants $\mu$ and $c$ are measures of the material's internal friction and cohesive strength, respectively. Stiffening requires the formation of a network of contact forces that constrain particle motions in all directions. As the applied stress increases, these contact forces must also increase, creating a subset of contact forces in the system that are much larger than those present due to gravity alone. We focus here on identifying the spatial form of this subset of particles responsible for the stiffening behavior.

\subsection{Yielding and stiffening}

\begin{figure}[tb]
 \centering
 \includegraphics[width=3.375 in,keepaspectratio=true]{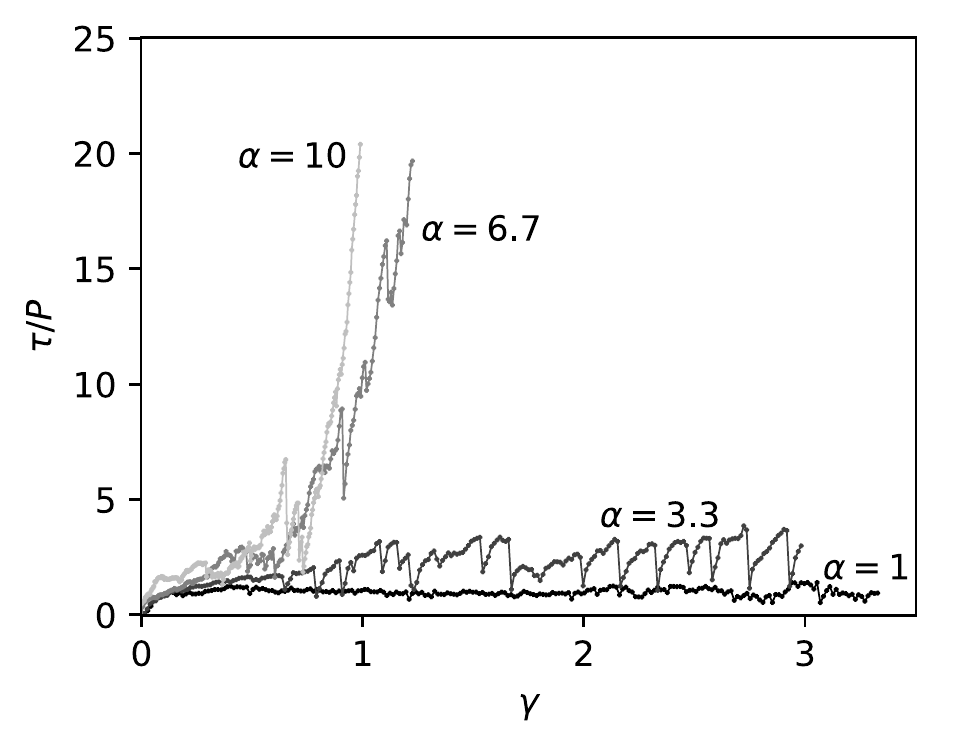}
 \caption{Typical shear response of packings with increasing particle $\alpha$ show a transition from yielding to stiffening. In the shown runs, $P=0.97\,$kPa ($\alpha=1$), $0.74\,$kPa ($\alpha=3.3$), $0.25\,$kPa ($\alpha=6.7$), and $0.2\,$kPa ($\alpha=10$)}.
 \label{fig:stiffening}
\end{figure}

Two alternative types of behavior are observed in individual runs: plastic yielding or stiffening, as demonstrated in Fig.~\ref{fig:stiffening}. Packings with small $\alpha$ deform plastically under shear: $\tau$ fluctuates about a steady-state value for $\gamma>1$, analogous to critical state in soil mechanics \cite{Wood1990_book1}. In contrast, for large $\alpha$, we observe a sharp increase in $\tau$ with increasing strain.

\begin{figure}[tb]
 \centering
 \includegraphics[width=3.375 in,keepaspectratio=true]{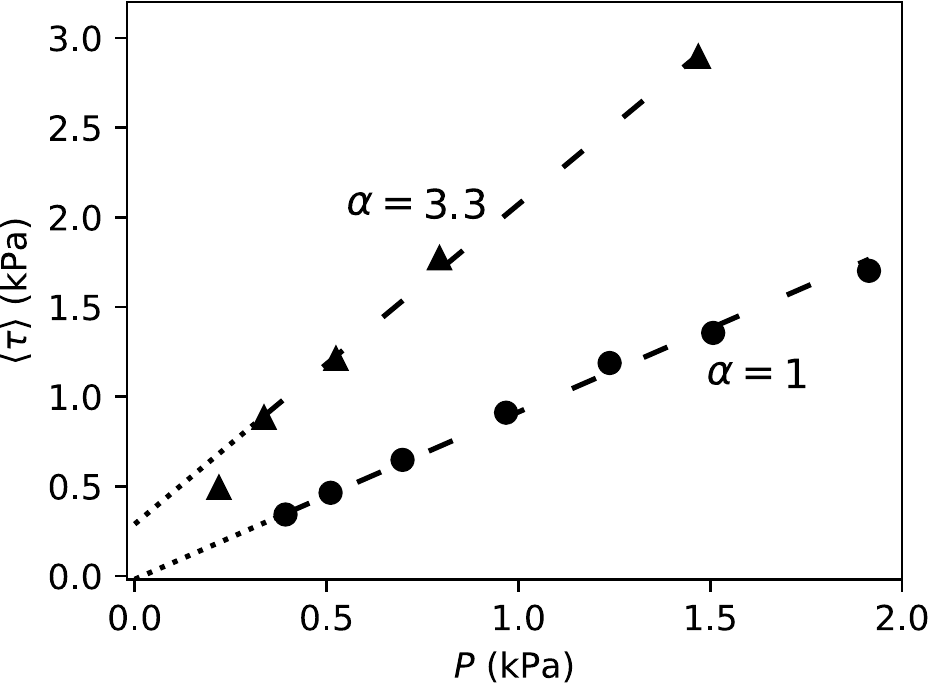}
 \caption{Yield points for packings of particles with $\alpha=1$ (circles) and $3.3$ (triangles), measured in experiments. The error bar is smaller than the marker size and thus is not shown. Dashed lines show linear least-squares fits, with slopes $0.93$ and $1.78$. Dotted lines show extrapolations to $P=0$.}
 \label{fig:mc1}
\end{figure}

For cases in which we observe yielding, a simple measure of $\mu$ and $c$ can be obtained by fitting the data for $\tau(P)$, as shown in Fig.~\ref{fig:mc1}. This corresponds to a standard procedure for characterizing systems in which the force from lateral boundaries can be neglected. For a given $P$, we average the shear stress $\tau$ for each run using only the data for $\gamma>1$ to avoid including the transient. We then average over different runs to get $\avg{\tau}$ and an estimate of the sample-to-sample fluctuations. The results are fit to the linear form $\avg{\tau}=\mu_{\rm expt} P+c_{\rm expt}$ using a least squares method. We find $\mu_{\rm expt} = 0.93 \pm 0.02$ and $c_{\rm expt}= 20 \pm 10\,$Pa for $\alpha = 1$ packings and $\mu_{\rm expt} = 1.78 \pm 0.06$ and $c_{\rm expt} = 290 \pm 50\,$Pa for $\alpha = 3.3$ with $68$\% confidence using standard methods. The fact that $\mu_{\rm expt}(\alpha=3.3)$ is greater than $\mu_{\rm expt}(\alpha=1)$ is consistent with other studies of sheared granular materials with anisotropic grain shapes \cite{azema2013_pre}. This is due to the increase of the effective friction caused by geometrical asperities of the particles. The $\mu_{\rm expt}(\alpha=3.3)$ packings also show larger fluctuations in $\tau/P$. We also note that $\mu_{\rm expt}(\alpha=1)=0.91$ corresponds to a friction angle $\tan^{-1}(\mu_{\rm expt})=42^{\circ}$, which is higher than the material's angle of repose $31\pm2^{\circ}$ (measured by tilting a box with an initially flat packing). This is due to the fact that $\tau$ and $P$, which are measured at the boundary, do not accurately estimate the normal and shear stress in the interior of the deforming material \cite{thornton2001_pg}.

The fits indicate that $c_{\rm expt}(\alpha=1)=0$, as expected, and also $c_{\rm expt}(\alpha=3.3)>0$, suggesting that there is a nonvanishing geometric cohesion effect for $\alpha = 3.3$. In our system, however, $P$ does not necessarily represent the normal stress at the shear plane because there may be significant vertical forces applied by the tube walls. We note also that the apparent cohesion for $\alpha=3.3$ appears to vanish for sufficiently low $P$. The lowest $P$ we can realize in our experiments, which is due to gravity alone and is not included in the above fit, has a lower $\avg{\tau}$ than the fitting trend (Fig.~\ref{fig:mc1}). Simulations presented below, where the pressure on the shear plane itself can be determined, suggest that there is actually no apparent cohesion for $\alpha = 3.3$.  The difference between the $\alpha=3.3$ and $\alpha=1$ experimental cases is traceable to the tendency of the former to sustain substantial downward forces from the walls, particularly from the bottom edge of the top tube.

For the stiffening packings, the strain corresponding to the onset of rapid stiffening fluctuates from run to run, presumably due to the packing preparation and the finite size of the system, which has a diameter of roughly six times the arm length of an $\alpha=10$ particle and contains 250 particles. For independent repeated runs with $\alpha = 6.7$ or $10$, the likelihood of stiffening was greater than $50$\% for both shapes at low $P$. Increasing $P$ or increasing the initial packing fraction by tapping produces stiffening in $100$\% of the trials. We have checked that this stiffening occurs in larger systems, both in additional experiments and in numerical simulations (see Sec.~\ref{sec:simu} and Fig.~\ref{fig:simuBench}).

\subsection{\label{sec:ctscanres} Packing structure in stiffening systems}

\begin{figure}[tb]
 \centering
 \includegraphics[width=3.375 in,keepaspectratio=true]{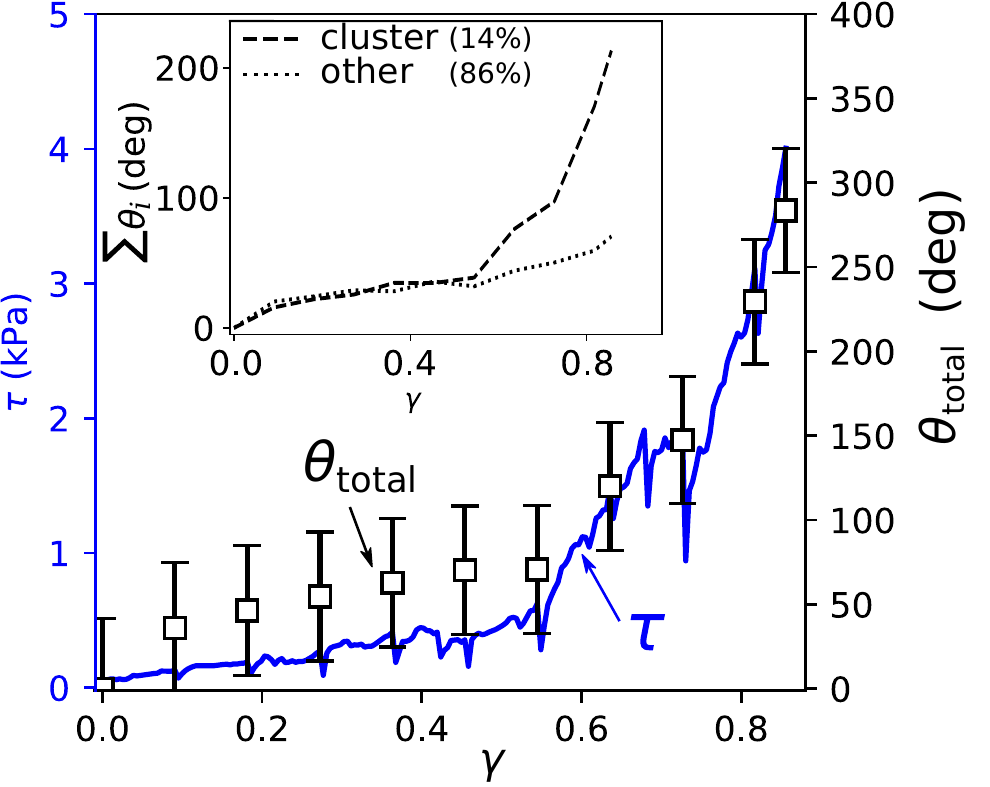}
 \caption{Total particle deformation $\theta_{\rm total}$ (open squares) and shear stress $\tau$ (solid line)\textit{vs.} strain $\gamma$ for $\alpha = 10$. The inset shows the particle deformation summed over the $14$\% of particles with the highest $\theta_i$ (dashed line) and the sum over the rest of the particles (dotted line).}
 \label{fig:bendStiffCorr}
\end{figure}

To identify key structures responsible for the stiffening, we first show that we can detect the bending of particle arms in the high stress states. Figure~\ref{fig:bendStiffCorr} shows that the x-ray CT protocol for identifying arm bending produces a signal that increases rapidly at approximately the same $\gamma$ where large forces develop. The open squares on the figure indicate $\theta_{\rm total} = \sum\theta_i$, where the sum is over all particles. Large contact forces do not necessarily result in substantial deformation of the particles because they may be applied close to the center of the particle. Nevertheless, the correlation between large applied force and the presence of bent arms is confirmed for independent runs. We find also that the particle deformation is concentrated in roughly $14$\% of the particles at each stage in the loading process. The inset in Fig.~\ref{fig:bendStiffCorr} shows that the particles in the top $14$\% of $\theta_i$ , selected at each strain independently, dominate the total deformation signal during stiffening. For convenience, we refer to these $14$\%\ as forming a \textit{rigid cluster} (C), and the rest of the particles as other particles (O). The results shown here and below are qualitatively similar for cutoff choices of $10$\% and $20$\%. During stiffening, the set of strongly deformed particles in C changes by less than 10\%, with fewer changes occurring during the later stages, indicating the emergence of a well-defined rigid cluster.  We note, however, that it is possible that the rigidity of the cluster C requires the presence of weak forces due to contacts with particles outside C.

\begin{figure}[tb]
 \centering
 \includegraphics[width=3.375 in,keepaspectratio=true]{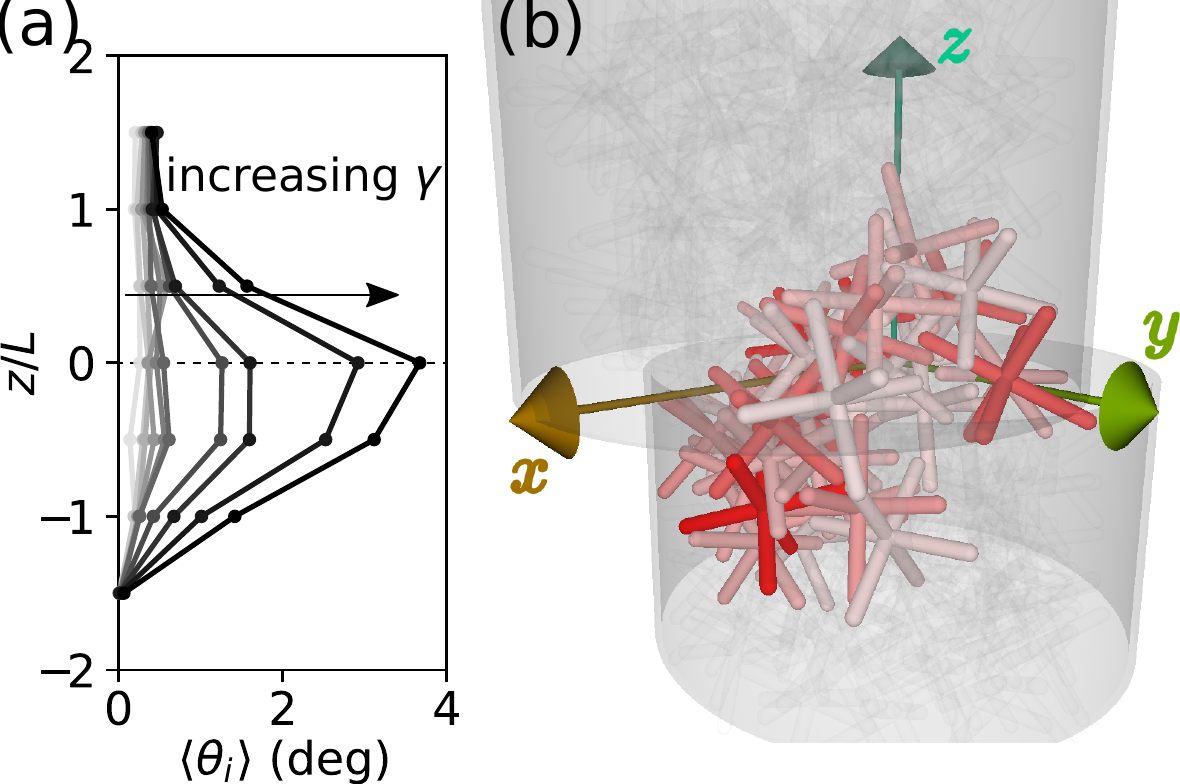}
 \caption{(Left) $\avg{\theta_i}$ as a function of height $z$ from the $z=0$ plane. Dark curve indicates large shear strain $\gamma$. (Right) Visualization of the packing at the last $\gamma$ in Fig.~\ref{fig:bendStiffCorr}. The $14$\% of particles with the highest $\theta_i$ are shown, with red indicating large $\theta_i$. The rest of the particles, which fill the tubes, are rendered semi-transparent.
 }
 \label{fig:bendLocalExpt}
\end{figure}

Figure~\ref{fig:bendLocalExpt} reveals that the most strongly bent particles are localized near the $z=0$ plane. We calculate the average $\theta_i$ for particles in a vertical window of width $0.5L$ centered at height $z$.  Before stiffening, $\avg{\theta_i} \approx 0$ for all $z$. During stiffening, $\avg{\theta_i}$ near $z=0$ increases dramatically, and greatly exceeds the original deformations due to gravity. A 3D rendering of the strongly deformed particles is shown in Fig.~\ref{fig:bendLocalExpt}, revealing a localization of the large stresses to a tilted band passing through the $z=0$ plane.

\begin{figure}[tb]
 \centering
 \includegraphics[width=3.375 in,keepaspectratio=true]{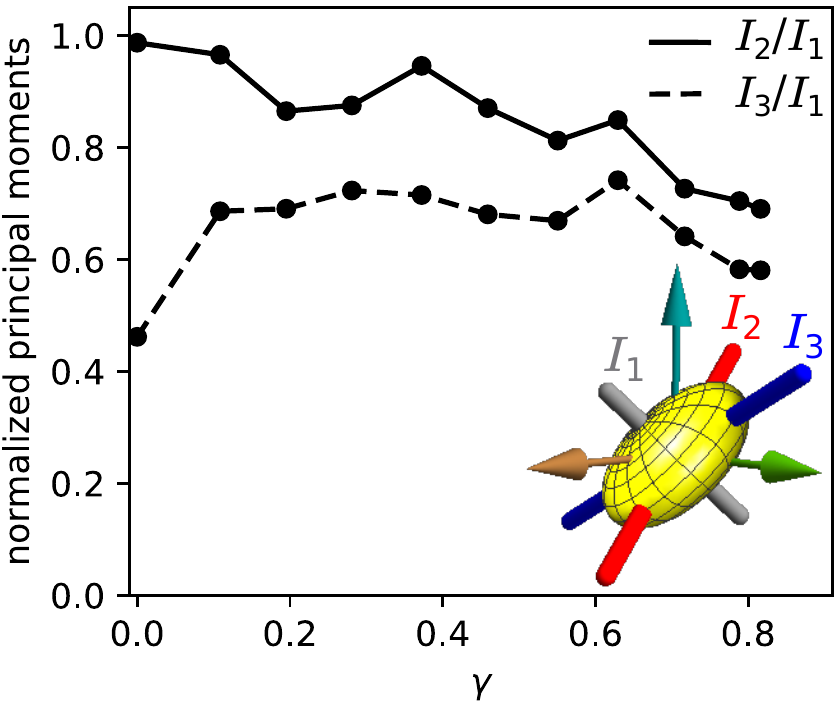}
 \caption{ Ratios of the principal moments of inertia, $I_2/I_1$ (solid line) and $I_3/I_1$ (dashed line), of the cluster $C$ as a function of shear strain in an experiment run.  The inset shows the approximated ellipsoidal shape of the cluster and its principal axes (rods). The 3D vectors indicate the lab coordinates, as in Figure~\ref{fig:bendLocalExpt}.}
 \label{fig:clusterShape}
\end{figure}

\begin{figure}[tb]
 \centering
 \includegraphics[width=3.375 in,keepaspectratio=true]{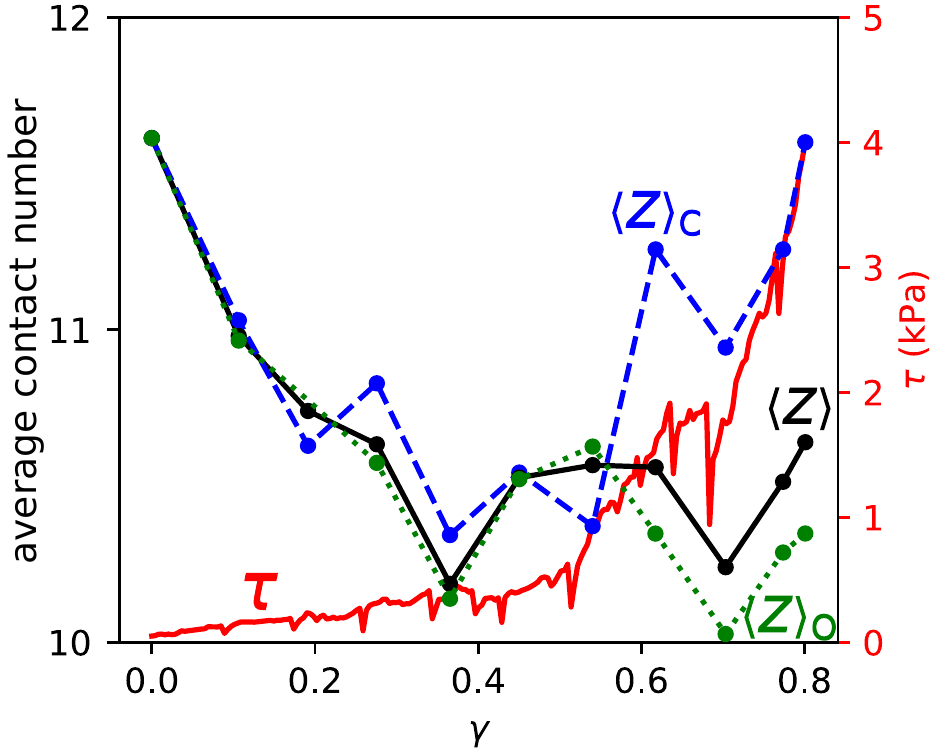}
 \caption{The evolution during a single run of the average contact number over all particles $\avg{Z}$ (black solid line), particles in the cluster $\avg{Z}_{\rm C}$ (blue dashed line), and the other particles $\avg{Z}_{\rm O}$ (green dotted line). The shear stress $\tau$ \textit{vs.} strain $\gamma$ (red line) indicates the stiffening.}
 \label{fig:contact}
\end{figure}

To characterize the shape of the stiffening cluster, we compute its principal moments of inertia $I_1$, $I_2$ and $I_3$ about its center of mass, with results as shown in Fig.~\ref{fig:clusterShape}. The ratios of intermediate and minor principal moments to the major moment are approximately $1.0$ and $0.45$ at $\gamma=0$ before stiffening, and $0.7$ and $0.6$ at $\gamma=0.86$ after stiffening, representing a change in cluster shape from a prolate to an oblate ellipsoid. 
(The original prolate shape simply represents a set of particles that are roughly uniformly distributed through the column.)
The contact network within this cluster evolves during stiffening, with new contacts being created. Figure~\ref{fig:contact} shows the average contact number for particles in the cluster, $\avg{Z}_{\rm C}$, along with a comparison to the average over the other particles $\avg{Z}_{\rm O}$ or over all particles $\avg{Z}$.
In the three experimental runs, the behavior of $\avg{Z}$, $\avg{Z}_{\rm C}$ and $\avg{Z}_{\rm O}$ varies substantially during the shearing phase, with all three decreasing in some runs and remaining constant in others.
The increase of $\avg{Z}_{\rm C}$ during stiffening is consistent across runs.
These added contacts within the cluster further increase its strength, leading to a strongly increasing shear modulus.

\section{\label{sec:simu}Numerical simulations}

We use the molecular dynamics software package LAMMPS \cite{plimpton1995_jcp} to simulate our direct shear experiments with particles having $\alpha=1$, $3.3$ and $10$. The equations of motion are integrated using the velocity Verlet scheme. Our simulation parameters and procedures are chosen to correspond reasonably well to our experiments, but there are features that we cannot match exactly. Most importantly, the particles in the simulations consist of rigidly connected spheres rather than smooth, flexible tubes. Arms are not allowed to bend, and the spacing between the spheres introduces geometric roughness that creates effective friction. Nevertheless, the simulations reproduce the main features of the experiments, suggesting that analyses of the detailed packing structures and forces within the simulations are indeed relevant for understanding the apparent cohesion and stiffening found by experiments.

\subsection{\label{sec:simuMD}Parameters and procedures}

\begin{figure}[tb]
 \centering
 \includegraphics[width=3.375 in,keepaspectratio=true]{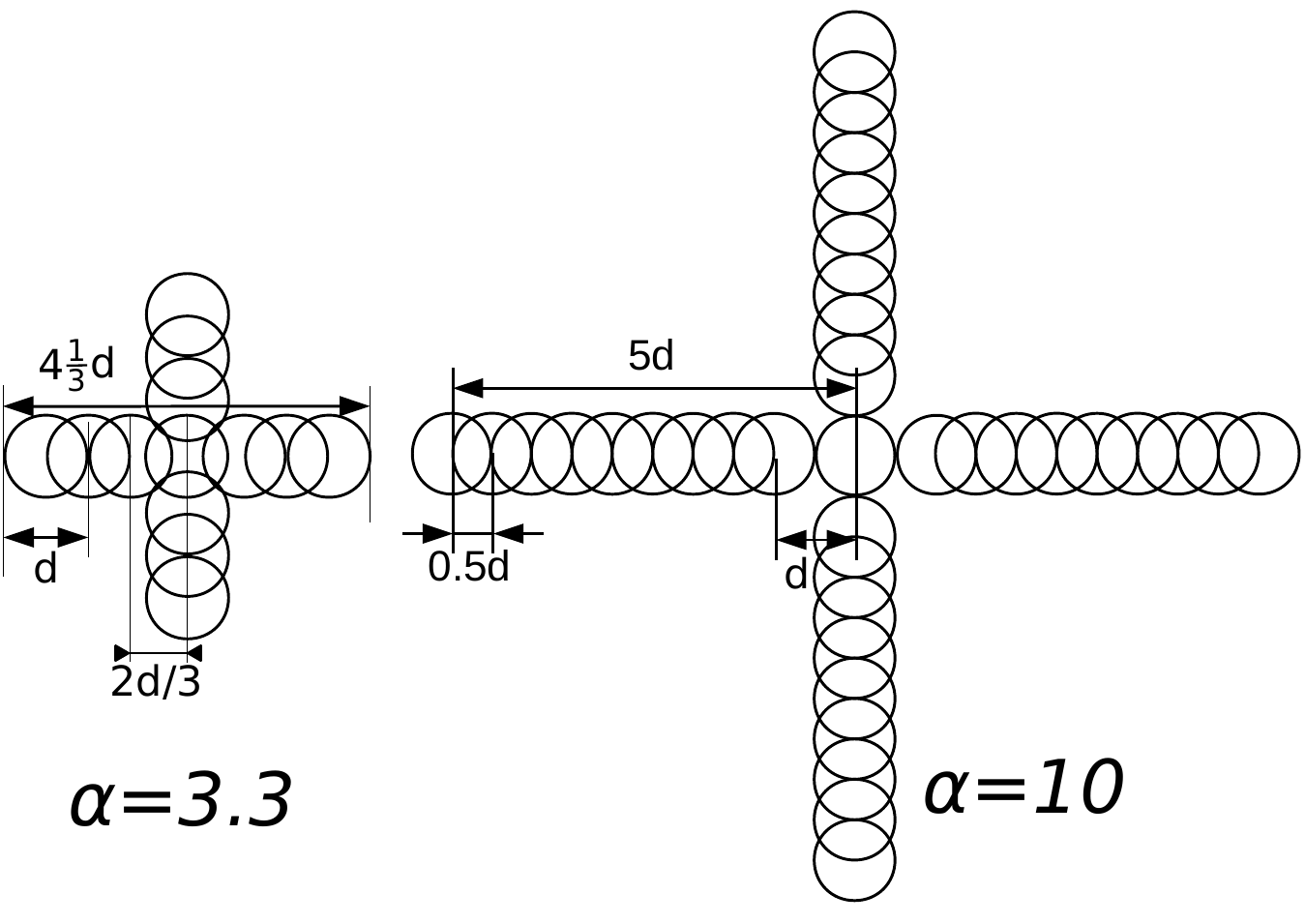}
 \caption{View of the two hexapods used in simulations, looking along the axis of one arm. Each circle is a finite-size sphere of diameter $d=3mm$ in the simulation.}
 \label{fig:hexapodTemplate}
\end{figure}

The particles sizes and shear cell dimensions in the simulations match those used in our experiments. The simulated $\alpha=3.3$ and $10$ hexapods are modeled as rigid bodies consisting of overlapping identical spheres, forming rough cylinders, as shown in Fig.~\ref{fig:hexapodTemplate}. The concavities in the arm surfaces create an effective friction coefficient equal to $0.27$ when two such concavities are nested within each other. All particle interactions are modeled as Hertz-Mindlin contact including Coulomb friction (using the pair\_style gran/hertz/history command in LAMMPS). In experiments, the material used to make $\alpha=1$ particles and $\alpha=3.3$ or $\alpha=10$ particles are different. To stay the same with experiments, we choose two corresponding sets of parameters for the contact model. The Young's modulus, Poisson ratio, and friction coefficient are $3\,$GPa, $0.35$ and $0.36$ for $\alpha=1$, and $1.5\,$GPa, $0.43$ and $0.3$ for $\alpha=3.3$ or $10$, respectively. The normal and tangential forces also contain damping terms linearly proportional to the relative velocity at the contact. The constant of proportionality is chosen so that the restitution coefficient for the collision of particle with a wall is close to experimentally measured value. For simplicity, we choose the normal and tangential damping coefficients to be the same. Varying the damping coefficient by an order of magnitude does not change the qualitative features of the yielding or stiffening responses.

Particle-tube interactions are modeled in the same way as particle-particle interactions. For contacts with the extended support, the frictional forces are set to zero to mimic the low friction associated with the Teflon sheet used in experiments, and, for technical reasons, normal forces are taken to be linearly proportional to the overlap distances between the particles and the extended support. Sample preparation and shear procedures in the simulations mimic the experiments. We randomly drop particles into the tube, releasing $n$ particles every $0.12$ seconds at random horizontal positions $18\,$cm above the bottom and letting them fall in place to create a packing. We take $n \approx 85$ for $\alpha = 1$ particles, $n = 115$ for $\alpha = 3.3$ particles, and $n = 5$ for $\alpha = 10$ particles. After the particles have settled under gravity, we shear the packing by displacing the top tube at a constant horizontal speed of $0.1\,$mm/s, up to a total displacement of $30\,$mm.

\subsection{\label{sec:simuRes}Results}
\begin{figure}[tb]
 \centering
 \includegraphics[width=3.375 in,keepaspectratio=true]{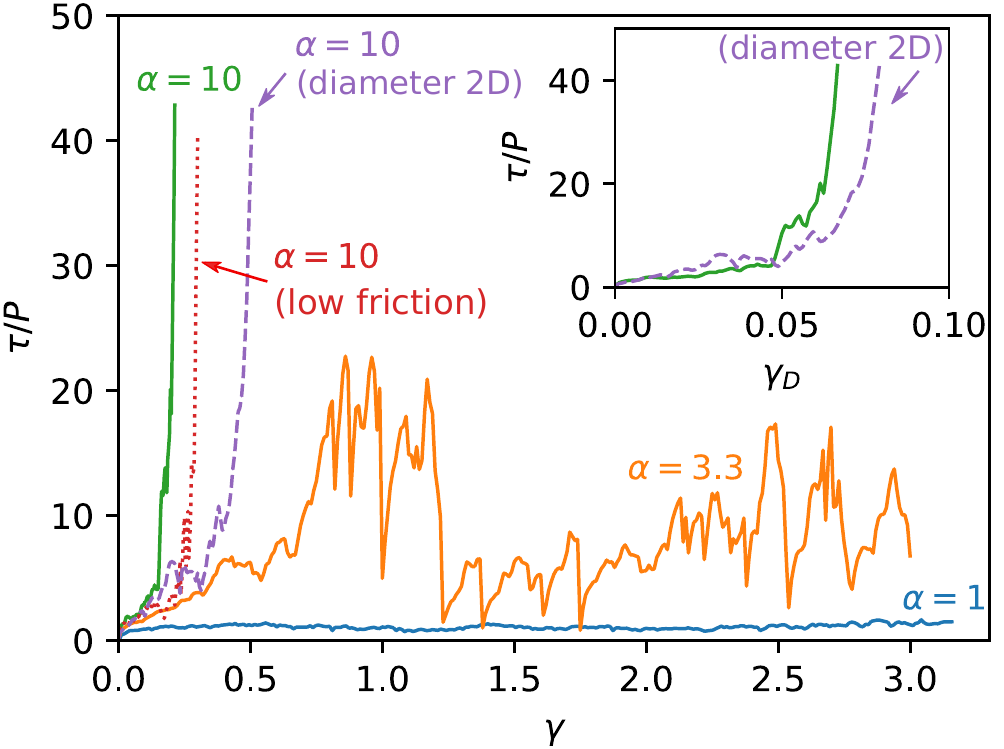}
 \caption{Ratio of shear stress to applied normal stress, $\tau/P$, \textit{vs.} strain $\gamma$ for numerical simulations of sheared packings with different particle shape aspect ratio $\alpha$. For $\alpha=10$, results are shown for two different tube diameters, $D$ and $2D$, with $D=96\,$mm, and low interparticle friction coefficient. $P=0.72\,$kPa,  $0.35\,$kPa, and $0.08\,$kPa for $\alpha=1$, 3.3, and 10, respectively, with $D=96\,$mm, and $P=0.068\,$kPa for $\alpha=10$ and tube diameter$=2D$. Inset: the same runs are plotted against $\gamma_D$, the ratio of horizontal tube displacement to tube diameter.}
 \label{fig:simuBench}
\end{figure}

Our simulations reproduce the qualitative plastic yielding and stiffening for the different particle $\alpha$, as shown in Fig.~\ref{fig:simuBench}. Quantitatively, the simulated materials appear to sustain stronger forces: for $\alpha=1$ or $3.3$, the shear strengths $\tau/P$ are greater than those in experiments; and for $\alpha = 10$, the transition to stiffening occurs at a smaller $\gamma$ ($\approx 0.2$) than in experiments ($\approx 0.8$). The latter effect is likely due to the increased effective interparticle friction created by the joined spheres that make up each arm. Specifying a smaller Coulomb friction coefficient between particles ($=0.1$) results in an increase in the strain for the onset of stiffening (Fig.~\ref{fig:simuBench}). Doubling the shear tube diameter, we find stiffening at roughly the same value of the strain, $\gamma_{D}$, defined as the ratio of horizontal tube displacement to the tube diameter (Fig.~\ref{fig:simuBench} inset).

\subsubsection{\label{sec:simuResAC}Apparent cohesion in yielding systems}

\begin{figure}[tb]
{\large (a)} \begin{minipage}[t]{0.9\columnwidth}
\vspace{0pt} 
\includegraphics[width=3.375 in,keepaspectratio=true]{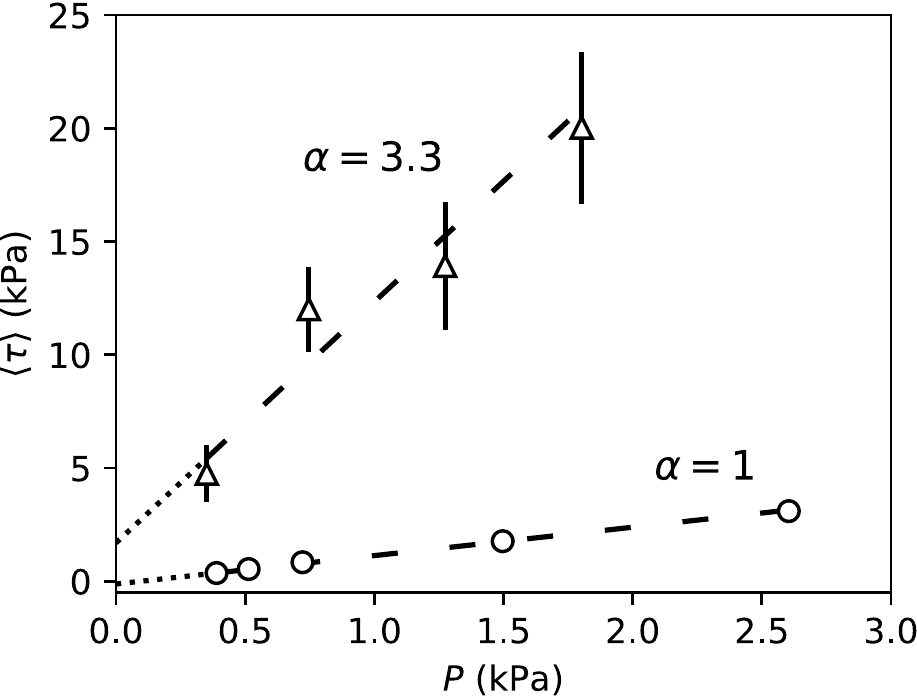}
\end{minipage} \\
{\large (b)} \begin{minipage}[t]{0.9\columnwidth}
\vspace{0pt} 
\includegraphics[width=3.375 in,keepaspectratio=true]{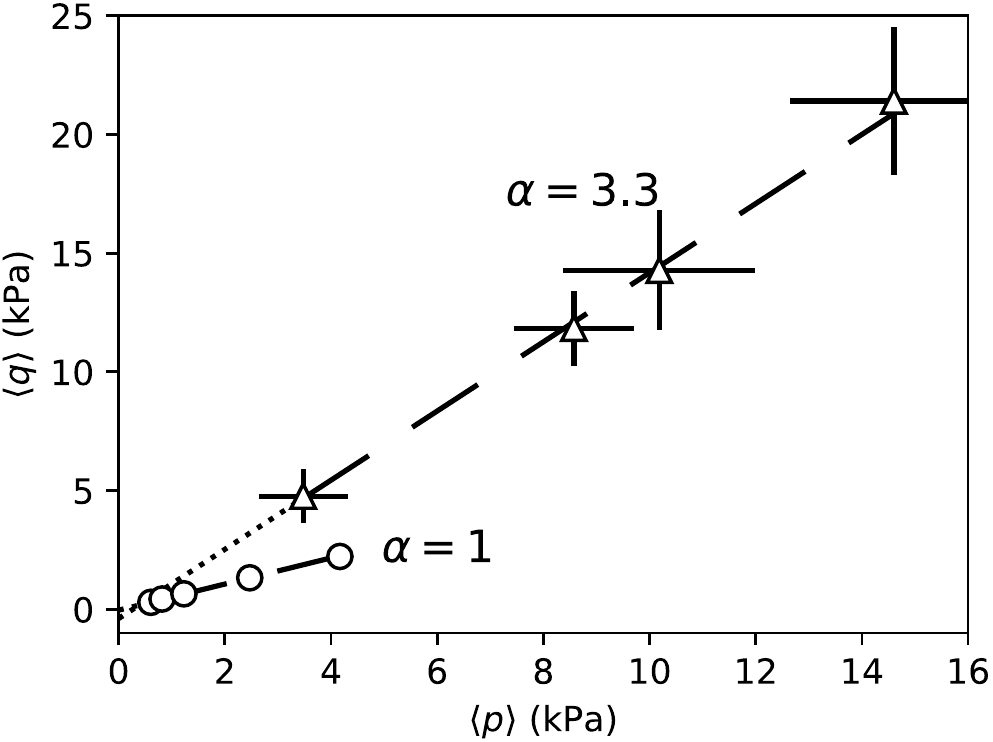}
\end{minipage}
 \caption{(a) The sample-averaged shear stress in the steady state as a function of applied normal stress for direct shear simulations with particle of aspect ratios $\alpha=1$ (circles) and $\alpha=3.3$ (triangles). The error estimates for $\alpha=1$ particle packings are smaller than the marker size. Dashed lines shown the linear least squares fits, with slopes $1.25$ and $11$~. Dotted lines show extrapolations to $P=0$. (b) Sample-averaged pressure $\avg{p}$ \textit{vs.} deviatoric stress $\avg{q}$ in the steady-state are shown for packings with particle $\alpha=1$ (open circles) and $3.3$ (triangles). Dashed lines shown the linear least squares fits, the dotted lines show extrapolations to $P=0$, with slopes $0.55$ and $1.5$~.}
 \label{fig:mc1simu}
\end{figure}

Figure~\ref{fig:mc1simu}(a) shows the sample-averaged shear stress in the steady state, $\avg{\tau}$, \textit{vs.} applied normal stress, $P$, for numerical simulations with particle aspect ratios $\alpha=1$ and $3.3$. The large error bars for the $\alpha = 3.3$ case are due to large fluctuations of the shear stress in the steady state in individual runs. We fit the results to the linear form $\avg{\tau} = \mu_{\rm sim} P + c_{\rm sim}$ using a least squares method. For $\alpha = 1$ packings, we find $\mu_{\rm sim} = 1.25 \pm 0.0$2 and $c_{\rm sim} = -0.12 \pm 0.01\,$kPa. For $\alpha = 3.3$, we find $\mu_{\rm sim} = 11 \pm 2$ and $c_{\rm sim} = 2 \pm 2\,$kPa. In both cases, the internal friction coefficient $\mu_{\rm sim}$ is larger than the value obtained from experiments. As expected, the $\alpha = 1$ case shows a vanishing apparent cohesion compared to the range of $P$, which is on the order of $1\,$kPa. Also as in experiments, $\alpha = 3.3$ particle packings show an apparent cohesion comparable to $P$. In this case, however, we will see that the apparent cohesion is an artifact, due to the fact that $P$ does not account for downward forces applied by the tube walls to the packing.

Using the pressure of the packings near the shear plane, which can be derived from the simulated contact forces, we find no apparent cohesion. The complete set of contact forces for a given snapshot of the simulation can be used to construct a stress tensor, $\bm\sigma$, associated with a single particle or collection of particles in a given region of the packing:
\begin{eqnarray}
\sigma=\frac{1}{V}\sum\limits_{i=1}^N\sum\limits_{k=1}^{Z_i} \mathbf{f}_{k,i} \otimes \mathbf{r}_{k,i}\,,
\label{eq:stressTensor}
\end{eqnarray}
where $N$ is the number of particles in a chosen volume $V$, which is a vertical window of width $4L$ centered at the $z=0$ plane. The second sum is over all of the contacts where forces are applied to particle $i$. The vector $\mathbf{r}_{k,i}$ points from particle $i$'s center to the point of contact, and $\mathbf{f}_{k,i}$ is the force on particle $i$. Given this definition, a negative (positive) principal stress means the material is under compression (tension). This definition of $\bm\sigma$ corresponds to the stress tensor computed based on forces on the boundary of volume $V$ \cite{rothenburg1989_geotec,moreau1997_friction}. From $\bm\sigma$, we calculate the pressure $p$ and deviatoric stress $q$, which are the responses to volumetric and distortional deformation of the material, defined as in Ref.~\cite{Wood1990_book1}:
\begin{equation}
p = \frac{1}{3}\mathrm{Tr}(\bm\sigma)\,;
\quad\quad
q = ||\frac{3}{2}(\bm\sigma-p\,\mathbf{I})||\,,
\label{eq:one}
\end{equation}
where $||\mathbf{a}||\equiv\sqrt{\sum_{i,j} a_{ij}a_{ij}}$.

Figure~\ref{fig:mc1simu}(b) shows the sample averaged $p$ \textit{vs.} $q$ for yielding systems. For a given $P$, we first average $p$ and $q$ for each run using only the data for $\gamma\geq1$ to avoid the transient. We then average over different runs to get $\avg{p}$ and $\avg{q}$, and estimate the sample-to-sample fluctuations. $\avg{p}$ and $\avg{q}$ for $\alpha=3.3$ packings show larger fluctuations than $\alpha=1$. The results are fit to the standard Dr\"{u}cker-Prager form of the yield condition \cite{andreotti2013_book} $\avg{q}=\mu_{\rm DP}\avg{ p}+c_{\rm DP}$, using the least squares method. Dashed lines in Fig.~\ref{fig:mc1simu}(b) show the fit, and dotted lines show extrapolation to $\avg{ p}=0$. Sample fluctuations for $\alpha=1$ particle packings are smaller than the marker size. As in experiments, we find that $\mu$ is larger for $\alpha=3.3$ packings than for $\alpha=1$. For $\alpha=1$ packings, we find $\mu_{\rm DP}=0.55\pm0.01$ and $c_{\rm DP}=-0.04\pm0.01\,$kPa. For $\alpha=3.3$, we find $\mu_{\rm DP}=1.5\pm0.4$ and $c_{\rm DP}=0\pm3\,$kPa. The apparent cohesion coefficient $c_{\rm DP}$ is consistent with zero for both particle types.

The discrepancy between $c_{\rm sim}$ based on the applied normal stress $P$ and $c_{\rm DP}$ extracted from those same simulations using the measured pressure $p$ is resolved by noting that large vertical forces are applied to the packing by the bottom edge of the top tube. Including contributions to $P$ from these forces gives a fit with $c_{\rm sim}=0$ within uncertainty (results not shown), consistent with the results obtained using $p$.  

\subsubsection{\label{sec:simuResCT}Packing structures in stiffening systems}

\begin{figure}[tb]
 \centering
 \includegraphics[width=3.375 in,keepaspectratio=true]{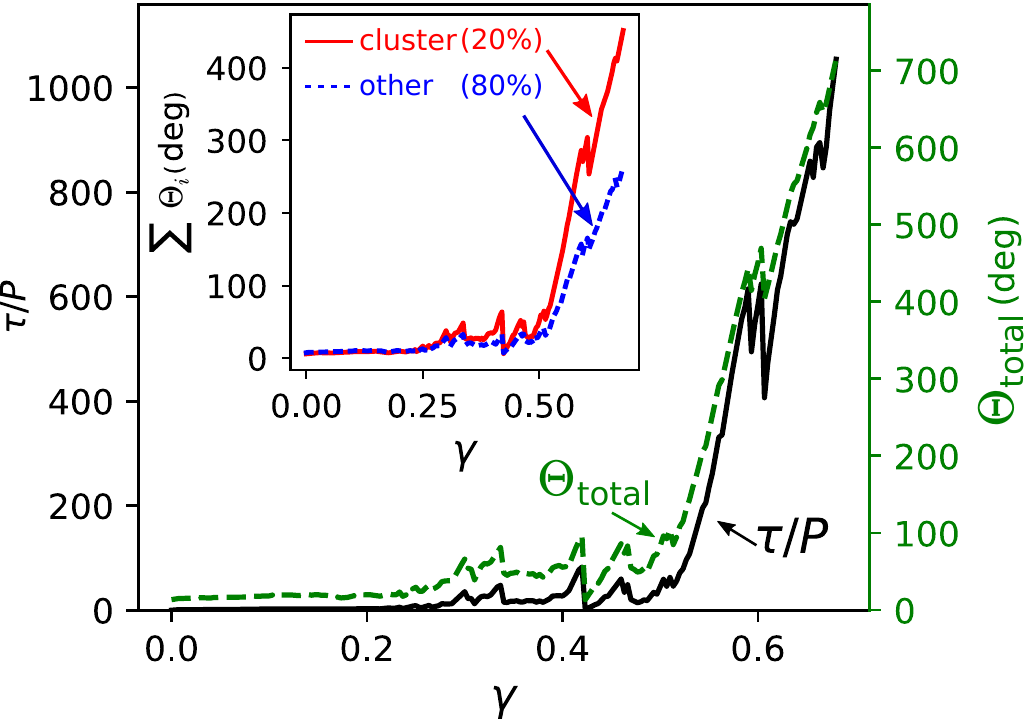}
 \caption{Total particle deformation $\Theta_{\rm total}$ (green dashed line) and shear stress divided by applied normal stress $\tau/P$ (black solid line) \textit{vs.} strain $\gamma$ for a simulation of $\alpha=10$ particle packing. The inset shows the particle deformation summed over the $20$\% of particles with the highest $\Theta_i$ (red solid line) and the sum over the rest of the particles (blue dashed line).}
 \label{fig:particleProf1}
\end{figure}

Our simulations for $\alpha=10$ particle packings reproduce the structures obtained from the CT measurements described in Sec.~\ref{sec:ctscanres}. Though the simulated particles are inflexible, the elastic energy is represented by allowing overlaps of arms, and the amount of interpenetration can be used as a proxy for the bending of arms in the experiment. For a given contact, we define $\delta$ as the overlap of two spheres on the contacting arms. We then define the quantity $\Theta_i$ as the sum of $\delta/\ell$ over all contacts of particle $i$, where $\ell$ is the moment arm length measured from the particle center. We take $\Theta_i$ to be the analog of the quantity $\theta_i$ measured from CT data. Figure~\ref{fig:particleProf1} shows that the sum over all particles $\Theta_{\rm total}=\sum\Theta_i$ is strongly correlated with the rapid increase in $\tau$. For the largest shear strains, $\Theta_{\rm total}$ is of the same order of magnitude as $\theta_{\rm total}$ (Fig.~\ref{fig:bendStiffCorr}). Note, however, that $\tau$ is much larger in simulations than in experiments.
This is because hexapod arms do not bend in simulations; the stress scale is set by the material stiffness (on the order of $10^{6}\,$N/m for an overlap of $3$\,mm) rather than the bending stiffness of arms (on the order of $10^{4}$\,N/m at arm tip), as in experiments.  The former was set to this high value to prevent arms from passing through each other for the relevant shear magnitudes.  (Recall that excessive force in the experiments leads to the breaking of particle arms.)

Consistent with the experiments, we find that $\Theta_{\rm total}$ is concentrated in roughly $20$\% of the hexapods. The inset in Fig.~\ref{fig:particleProf1} shows that $\sum\Theta_i$ over the particles in the top $20$\%, selected at each strain independently, increases roughly twice as fast as the sum over the lower $80$\%, indicating that these top $20$\% are primarily responsible for the stiffening of the system. For convenience, we refer to these $20$\% of particles as a \textit{rigid cluster} (C), and the remaining set ``others'' (O). The results here and below are qualitatively similar for a cutoff choice anywhere between $16$\% and $24$\%.  

\begin{figure}[tb]
 \centering
 \includegraphics[width=3.0 in,keepaspectratio=true]{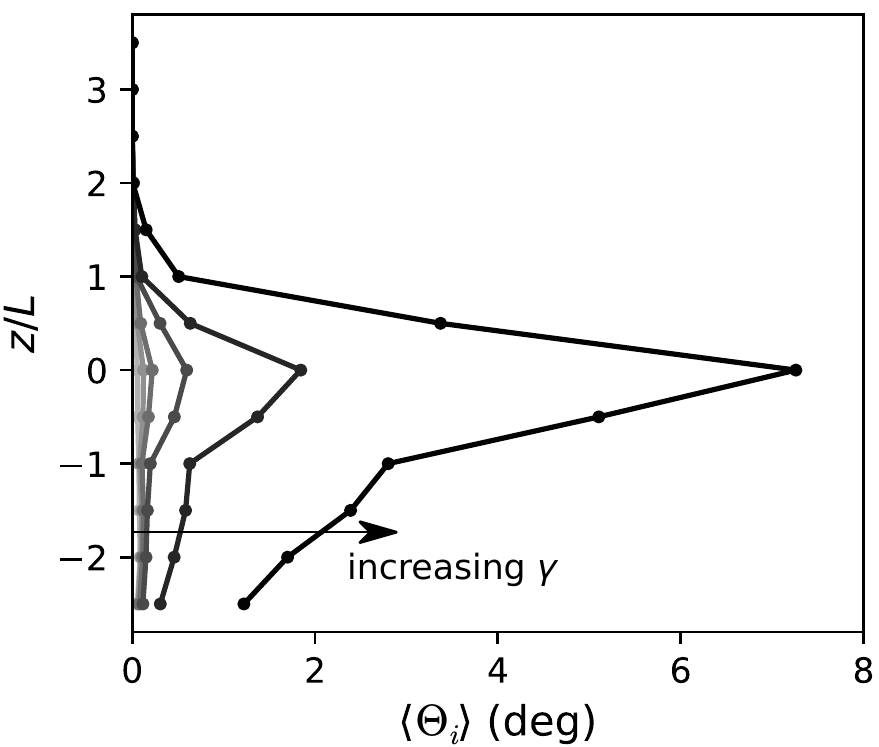}
 \caption{Sum of arm deformation on each particle, $\Theta_i$, averaged over particles whose centers lie in a horizontal window centered at $z$.  Darker curves indicate larger shear strain.}
 \label{fig:particleProf2}
\end{figure}

As expected, the stress associated with stiffening is localized near the nominal shear plane. Figure~\ref{fig:particleProf2} shows the average $\Theta_i$ for particles in a vertical window of width equal to $L/2$, centered at a height $z$. The stiffening response indicated by $\avg{\Theta_i}$ is localized near the $z=0$ plane. As in the experiments (see Fig.~\ref{fig:bendStiffCorr}), during stiffening $\avg{\Theta_i}$ near $z=0$ increases dramatically. 
We stop the simulation when the overlap of particles is roughly 10\% of the arm diameter.  Increasing the applied force indefinitely would result in particle arms passing through each other, which is an irrelevant regime for the interpretation of experiments.

\begin{figure}[tb]
 \centering
 \includegraphics[width=3.375 in,keepaspectratio=true]{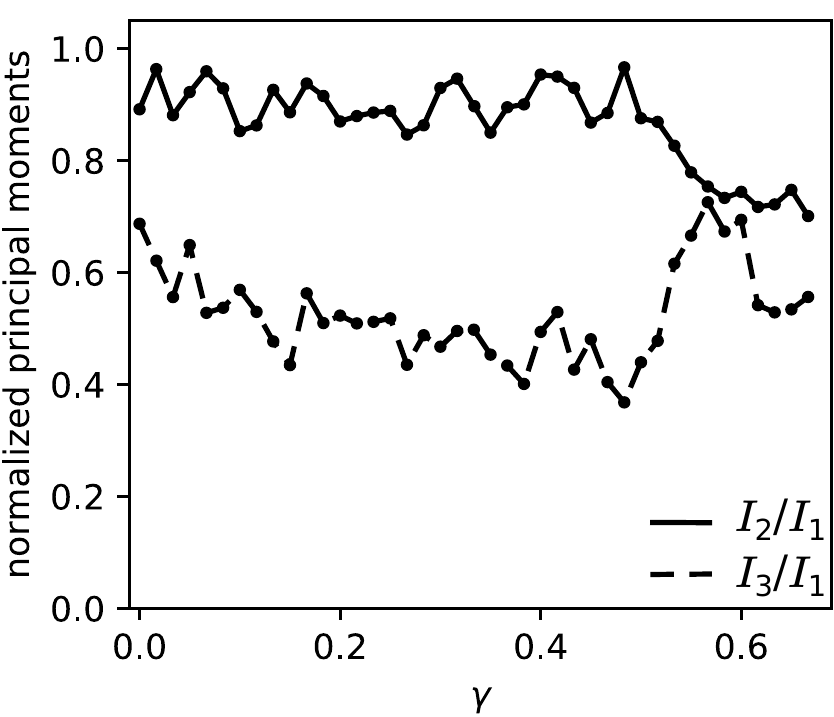}
 \caption{Ratios of the principal moments of inertia, $I_2/I_1$ (solid line) and $I_3/I_1$ (dashed line), of the cluster $C$ as a function of shear strain in a simulation run.}
 \label{fig:clusterShapeSimu}
\end{figure}

The method described in Sec.~\ref{sec:ctscanres} is used to characterize the shape of the stiffening cluster. As shown in Fig.~\ref{fig:clusterShapeSimu}, the ratios of the principal moments of inertia to the largest moment are approximately $0.5$ and $0.9$ before stiffening ($\gamma\approx0.5$), and $0.6$ and $0.7$ after stiffening, representing a change in cluster shape from a prolate to an oblate ellipsoid, consistent with the experimental results (Fig.~\ref{fig:clusterShape}).

\begin{figure}[tb]
 \centering
 \includegraphics[width=3.375 in,keepaspectratio=true]{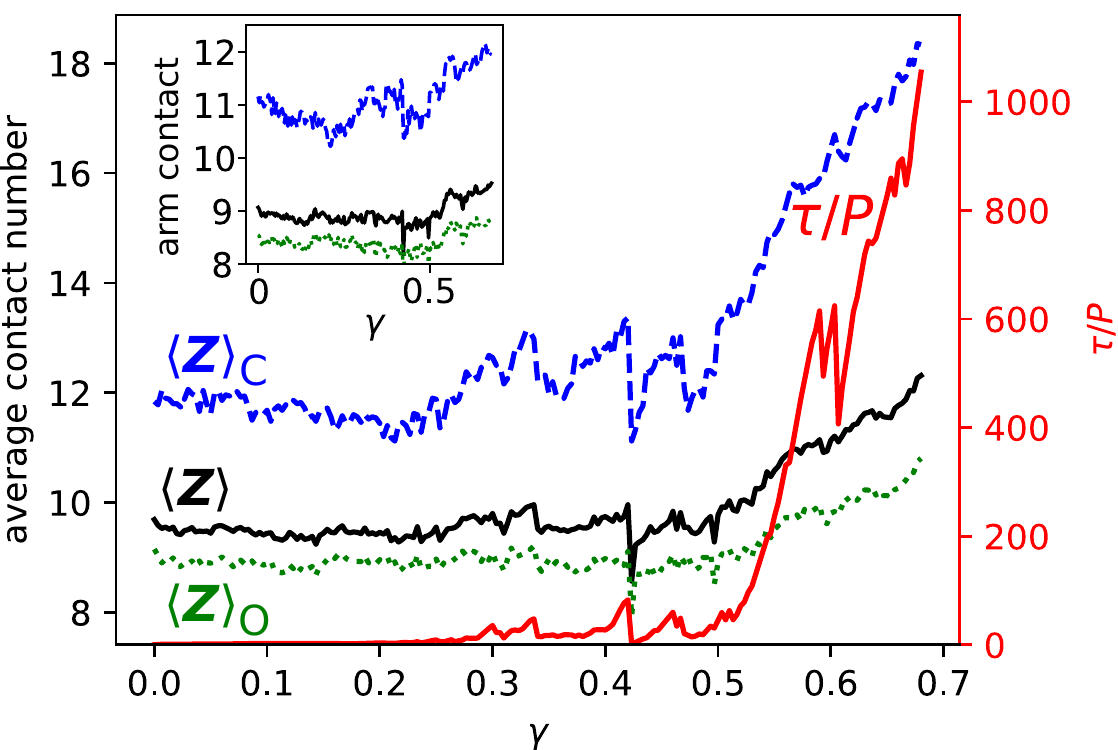}
 \caption{The evolution of the average contact number over all particles $\avg{ Z}$ (black solid line), particles in the cluster $\avg{ Z}_{\rm C}$ (blue dashed line), and the others $\avg{ Z}_{\rm O}$ (green dotted line) as a function of shear strain $\gamma$, from a numerical simulation of $\alpha=10$ particle packing under shear. The red solid line shows shear stress divided by applied normal stress $\tau/P$ \textit{vs.} $\gamma$. Inset: the evolution of $\avg{Z}$, $\avg{Z}_{\rm C}$ and $\avg{Z}_{\rm O}$ when a given pair of contacting arms is always counted as a single contact. Colors and line styles match the main figure.}
 \label{fig:simuContactNumber}
\end{figure}

Finally, Fig.~\ref{fig:simuContactNumber} shows the behavior of the average contact number during stiffening. $\avg{Z}$, $\avg{Z}_{\rm C}$, and $\avg{Z}_{\rm O}$ denote averages over all particles, over the rigid cluster, and over the others. We see that $\avg{Z}_{\rm C}$ increases substantially faster than $\avg{Z}_{\rm O}$, as in the experiments (Fig.~\ref{fig:contact}). Note that in our simulations a sphere on one arm can create contacts with two neighboring spheres on another arm, whereas in experiments a given pair of arms can have only one contact. Counting the number of arm contacts rather than sphere contacts reduces the $\avg{Z}$ values but does not change the relative trends during stiffening, as shown in the inset of Fig.~\ref{fig:simuContactNumber}. Thus it appears that during stiffening the rigid cluster strengthens by adding new contacts between particles.

\subsubsection{\label{sec:simuResST}Identification of tensile stresses}

A typical material with a positive Poisson ratio must support internal tensile stresses when subjected to external uniaxial compressive or shear forces.  In uniaxial compression of a cylinder, for example, tensile stresses must arise to counteract the tendency of the cylinder to bulge in the middle.
Similarly, for our granular packings, which tend to dilate through the top free surface (in the direction perpendicular to the shear plane), rigidity requires that there be some counterbalancing mechanism providing a tensile contribution in the packing. In this section, we identify a region in the simulated packing that is under tensile stress, and we elucidate the mechanism for supporting tensile stresses at the particle level. 

\begin{figure}[tb]
 \centering
 \includegraphics[width=0.9\columnwidth,keepaspectratio=true]{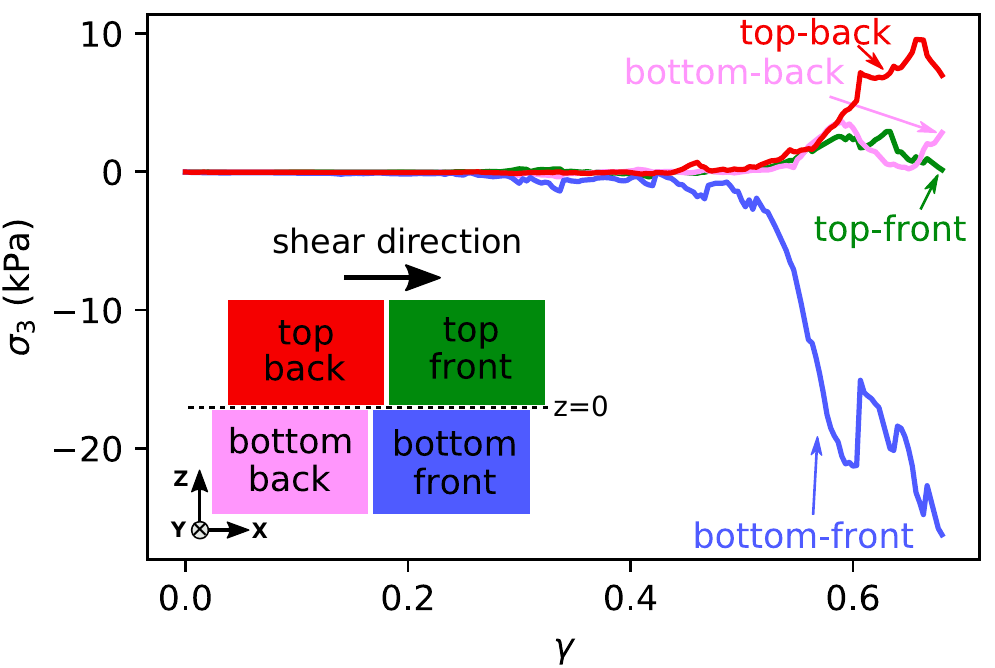}
 \caption{The minor principal stresses $\sigma_3$ \textit{vs.} strain $\gamma$ for $\alpha=10$ particles in the four different regions defined in the text. Positive values indicate tensile stress. Inset: Schematic showing the four regions used to calculate the average stress tensors.}
 \label{fig:tensileSimu}
\end{figure}

We first consider the average stresses within the four equal volume regions shown in the Fig.\ref{fig:tensileSimu} inset, which shows that there are significant variations within the packing. Each region is a semicircular portion of a cylinder with height $L$, covering the portion of the packing where highly stressed particles are found at large strains. The two regions above the $z=0$ plane are moving with the top tube, and the two bottom regions are fixed. For each region, we compute the stress tensor using Eq.~(\ref{eq:stressTensor}), averaging over particles whose centers lie within the region. The major principal stress $\sigma_1$ is compressive everywhere and is substantially stronger in the top-back and bottom-front regions, where it is oriented roughly in the $x$-$z$ plane, at a small angle to the $x$ axis. The intermediate principal stress is also compressive and is oriented along the $y$ axis. The minor principal stress $\sigma_3$ is compressive in the bottom-front region, but tensile in the top-back region, as shown in Fig.\ref{fig:tensileSimu}. In both cases, it is oriented roughly in the $x$-$z$ plane and close to the $z$ axis.

\begin{figure}[tb]
 \centering
 \includegraphics[width=3.375 in,keepaspectratio=true]{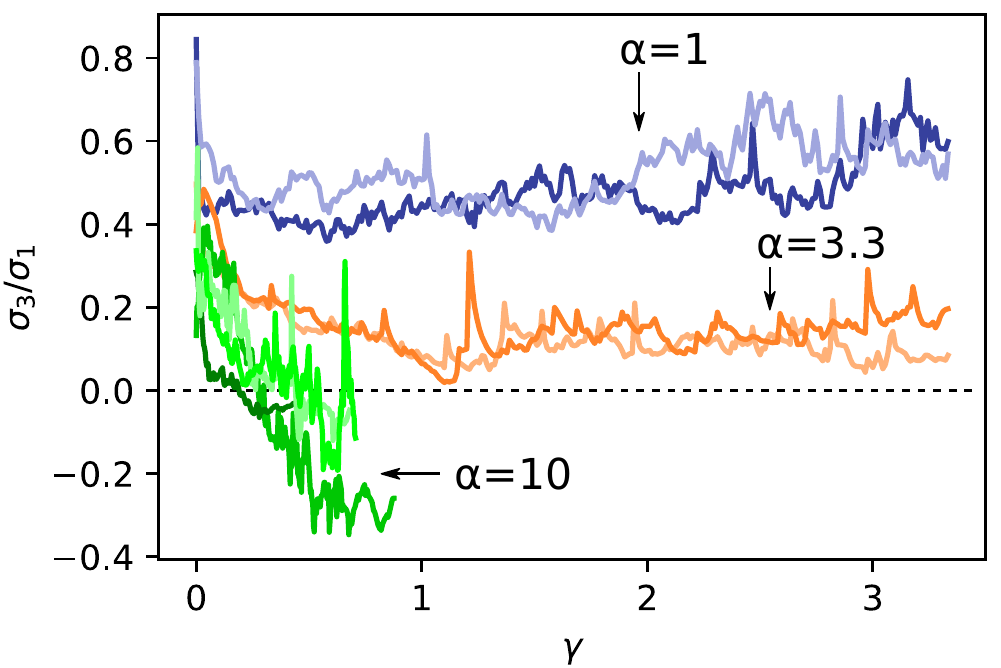}
 \caption{Ratios of minor to major principal stress $\sigma_3/\sigma_1$ in the top-back region (Fig.~\ref{fig:tensileSimu}) are calculated for different $\alpha$s and runs, and plotted as a function of shear strain $\gamma$. Colors refer to different $\alpha$, and lightness of the colors refer to different runs.}
 \label{fig:tensileRatio}
\end{figure}

The packing is not constrained externally from dilating in the positive $z$ direction. The tensile stresses with large projections on the $z$ axis resist the dilation that occurs in  packings with small $\alpha$ and allows them to flow instead of stiffen. Figure~\ref{fig:tensileRatio} shows $\sigma_3/\sigma_1$ in the top-back region as a function of $\gamma$ for several runs for particles with $\alpha=1$, $3.3$, and $10$. As $\alpha$ is increased, $\sigma_3/\sigma_1$ decreases faster with $\gamma$. $\sigma_3/\sigma_1$ remains positive during the shear for packings that yield, but it goes negative for the stiffening $\alpha=10$ packing at relatively small $\gamma$. Because $\sigma_1$ is always compressive, a negative ratio indicates that $\sigma_3$ is tensile. Thus we see that the change from yielding to stiffening behavior is correlated with the ability of the packing to support tensile stress.

\begin{figure}[tb]
 \centering
 \includegraphics[width=3.375 in,keepaspectratio=true]{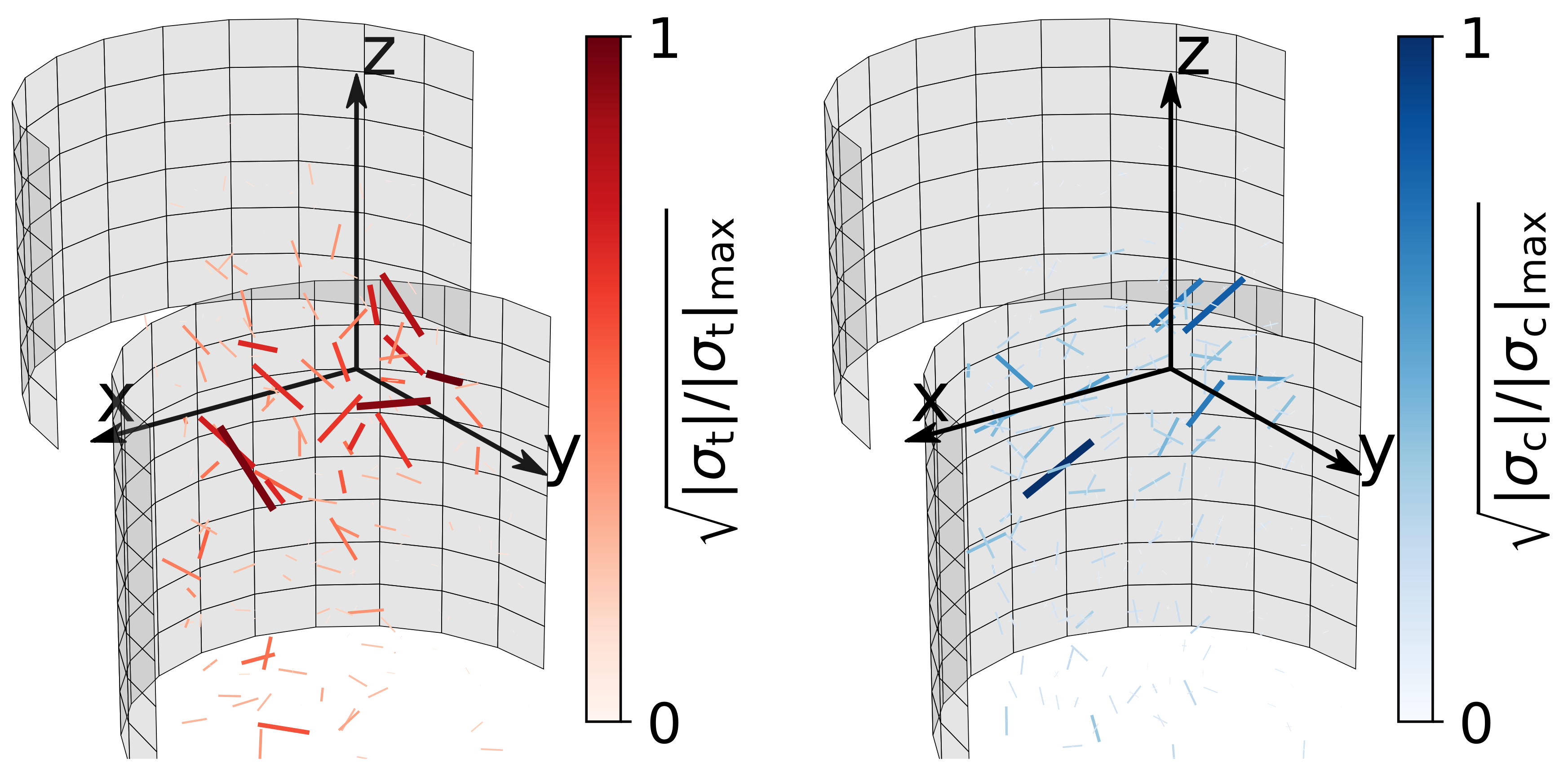}
 \caption{Visualization of principal stresses for $\alpha=10$ packing during stiffening. The principal stresses of each hexapod are calculated and drawn as line segments centered at the particle's center of mass. Tensile (compressive) principal stresses are shown on the left (right) in red (blue). The line color, length and thickness all vary linearly with the square root of stress magnitude and are normalized to the maximum stress magnitude. Gray grid surfaces represent the shear tube boundaries.}
 \label{fig:stressFieldVis}
\end{figure}

Figure~\ref{fig:stressFieldVis} shows principal stresses at the particle scale, calculated from Eq.~(\ref{eq:stressTensor}) by summing over the contacts of individual particles. Compressive and tensile principal stresses are represented as line segments centered at the particle's center. The line darkness, length, and thickness all vary linearly, with values normalized to the maximum magnitude in each panel. Compressive principal stresses $\sigma_{\rm c}$ are shown on the right (in blue). Tensile principal stresses $\sigma_{\rm t}$ are shown separately on the left (in red), as they would be difficult to see if normalized on the same scale as $\sigma_{\rm c}$. The rigid cluster is discernible in this figure as the collection of highly stressed particles. The large compressive stresses tend to align along the direction of the major axis of the rigid cluster ellipsoid, and the tensile stresses are close to the minor axis.

\begin{figure}[tb]
 \centering
 \includegraphics[width=1 in,keepaspectratio=true]{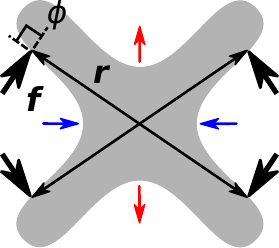}
 \caption{A cross-shape particle is under four contact forces $\mathbf{f}$ (big thick arrows) and the vectors $\mathbf{r}$ (small thin arrows). $\phi$ is defined as the acute angle $\phi$ between $\mathbf{f}$ and $\mathbf{r}$. In this case, $\phi=90^{\circ}$ for all contacts. Horizontal (blue) and vertical (red) arrows show the compressive and tensile principal stresses carried by the cross.}
 \label{fig:tensileExp}
\end{figure}

The source of tensile stress on a single particle can be understood as follows. In $\alpha=10$ packings, contacts between particle arms tend to occur far from the particle centers~\cite{bares2017_pg}, and the angle $\phi$ between the contact force $\mathbf{f}$ and the vector pointing from particle center to the contact $\mathbf{r}$ is expected to be nearly $90^{\circ}$, which implies that the contact force exerts a substantial torque on each particle~\cite{Murphy2016_gm}. To see that such forces can give rise to tensile stresses, consider a cross-shaped rigid particle subjected to four equal magnitude contact forces with $\phi=90^{\circ}$ as shown in Fig.~\ref{fig:tensileExp}. The configuration is in mechanical balance, and the stress tensor of Eq.~\ref{eq:stressTensor} has the form:
\begin{eqnarray*}
\sigma=
\left(
\begin{array}{cc}
-\sigma_0 & 0\\
0 & \sigma_0
\end{array}\right)\,,
\label{eq:3}
\end{eqnarray*}
indicating a tensile principal stress in the vertical direction and a compressive principal stress in the horizontal direction.

\begin{figure}[tb]
 \centering
 \includegraphics[width=3.375 in,keepaspectratio=true]{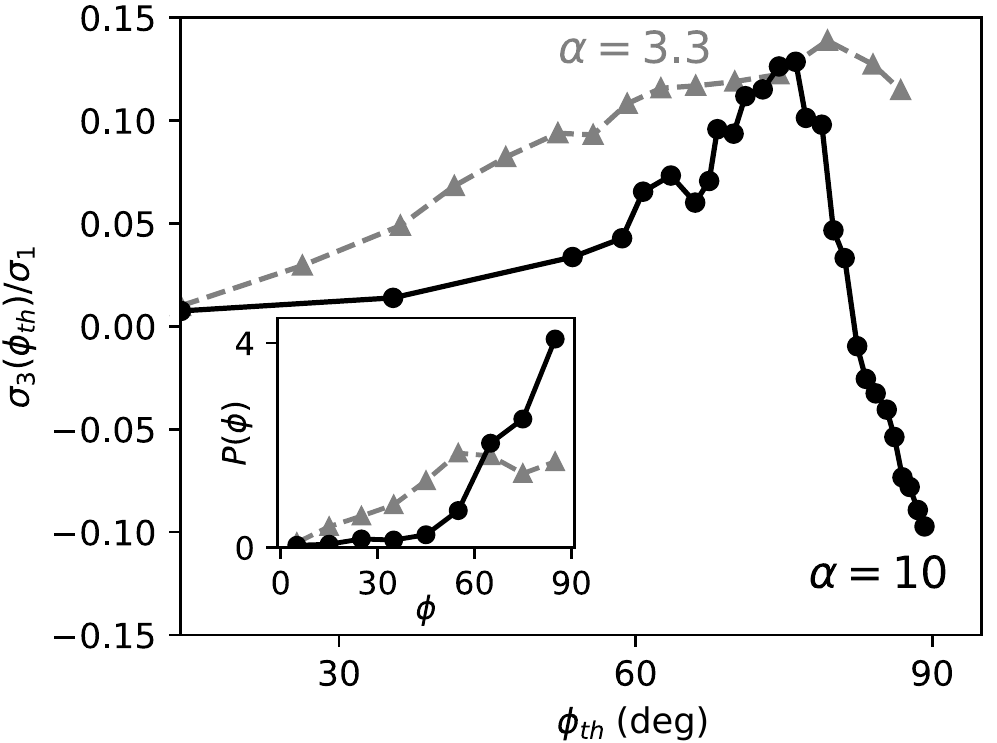}
 \caption{The minor principal stress $\sigma_3$ due to contacts that have an angle $\phi$ (defined in the text) less than a threshold $\phi_{th}$, $\sigma_3(\phi_{th})$, is normalized by major principal stress $\sigma_1$ and plotted \textit{vs.} $\phi_{th}$ for $\alpha=3.3$ (gray dashed line) and $10$ (black solid line). Inset: the distribution (arbitrary unit) of $\phi$ for the two $\alpha$s.}
 \label{fig:decomposition}
\end{figure}

The observation that contact forces with large $\phi$ (near $90^{\circ}$) are responsible for the tensile stress on individual particles implies that these contact forces are also responsible for the macroscopic tensile stress. Fig.~\ref{fig:decomposition} emphasizes this point by showing the net contribution to the global $\sigma_3$ from all contact forces with $\phi<\phi_{\rm th}$ as a function of $\phi_{\rm th}$. For the $\alpha=10$ case, we see that $\sigma_3/\sigma_1$ becomes negative only when contacts with $\phi \gtrsim 80^{\circ}$ are included. In contrast, for the $\alpha=3.3$ case, $\sigma_3(\phi_{th})/\sigma_1$ the contributions from contacts with large $\phi$ are not sufficient to generate a global tensile stress. This may be because there are fewer contacts with large $\phi$  (Fig.~\ref{fig:decomposition} inset). Alternatively, it could be that contacts with large $\phi$ do generate large tensile stresses on individual particles, but these are not well enough aligned to yield a net collective effect.

The tensile stress in these packings differ from those arising in cohesive granular materials like wet sand. As indicated by Fig.~\ref{fig:tensileExp}, the tensile stress in the hexapod packings is induced by applied compressive stress in an orthogonal direction. In the absence of compressive stresses, the system cannot support tensile stress. This is consistent with experimental observations of yield stress made for three-point bending tests on columns of Z-shape particles, in which the yield stress increased when the axial confining pressure of the column was increased \cite{Murphy2016_gm}.  As there is no other relevant quantity with dimensions of stress, the magnitude of the tensile stress must scale with the applied compressive stress.

\section{\label{sec:conclusion}CONCLUSIONS}

We perform experiments and simulations to analyze sheared granular materials with hexapod particles of increasingly non-convex shape, and we observe the development of structural rigidity when the arm length-to-diameter aspect ratio $\alpha$ is sufficiently large. For moderate aspect ratio ($\alpha=3.3$), the packings have a yield stress that vanishes for vanishing pressures, as in the case of hard spheres, suggesting that there is no effective cohesion in these systems (Fig.~\ref{fig:mc1simu}).

For packings that stiffen under direct shear ($\alpha=10$), x-ray Micro-CT data reveals that particle arms bend significantly, which allows for the identification of a cluster of particles responsible for the rigidity. We find that the stress is carried by an oblate cluster of particles localized near the nominal shear plane, and tilted slightly with respect to the plane (Fig.~\ref{fig:clusterShape}). 
The average contact number of the particles within the rigid cluster increases faster than that for particles outside the cluster, suggesting that the rigidity is due to the emergence of a collectively interlocked cluster, even though pairwise interactions of particles cannot act as hooks that support tensile stress.

Our numerical simulations reproduce the main features of the experiments and provide insights into the mechanism that leads to stiffening. Individual particles support tensile stresses arising from contact forces that are nearly perpendicular to the particle's arm, and such tensile stresses are organized so as to provide a macroscopic tensile stress in regions of the sheared random packing (Fig.~\ref{fig:decomposition}). This tensile stress can prevent dilation, allowing a cluster of particles to stabilize the packing against shear, and has magnitude proportional to the compressive stress acting in the orthogonal direction. New contacts are formed within the cluster as the strain is increased, leading to increasing stiffness.  

\begin{figure}[tb]
 \centering
 \includegraphics[width=3.3 in,keepaspectratio=true]{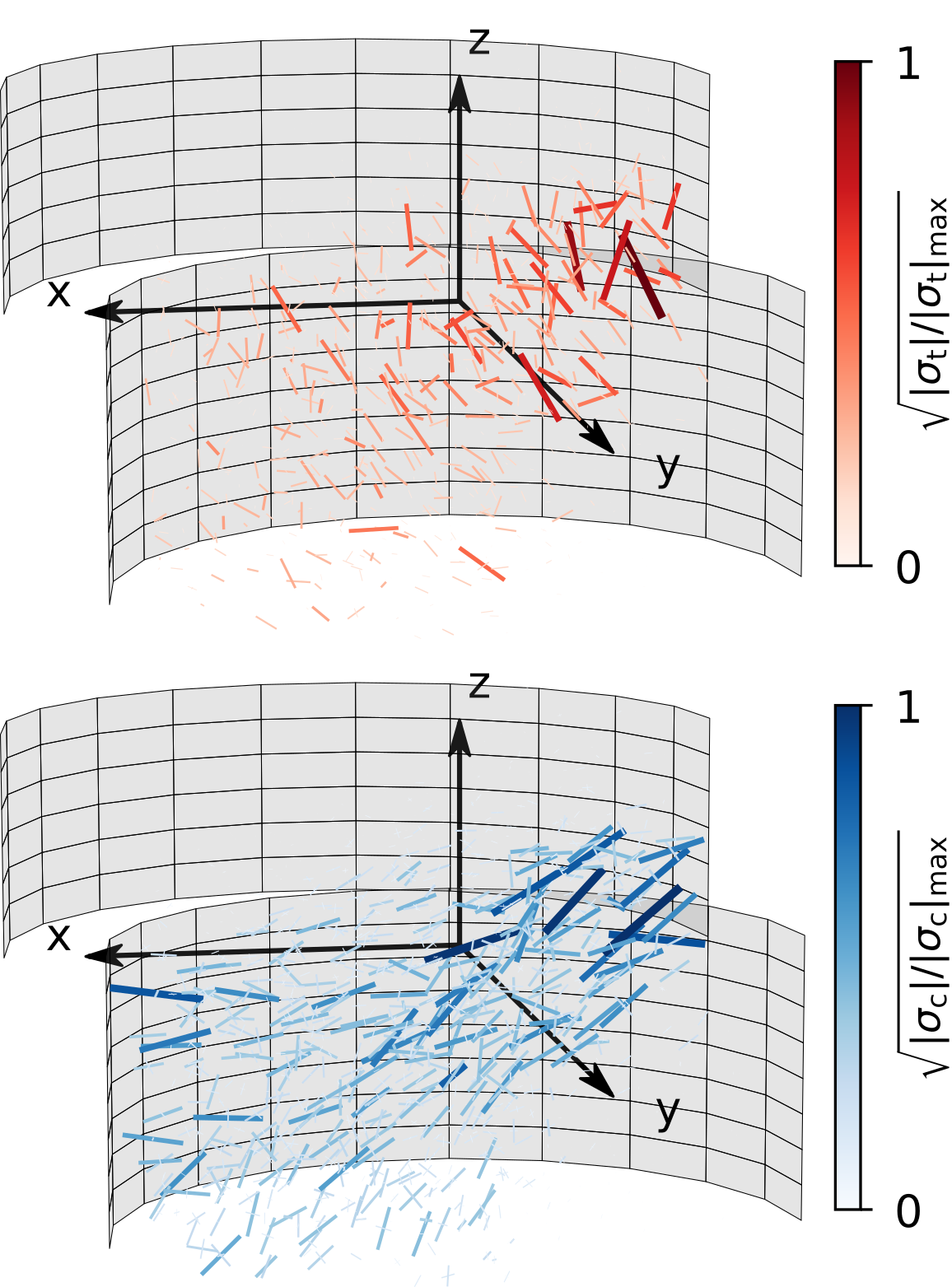}
 \caption{Visualization of tensile (top) and compressive (bottom) principal stresses for $\alpha=10$ packing in a tube of diameter twice as large during stiffening. The illustration methods are the same as in Fig.~\ref{fig:stressFieldVis}.}
 \label{fig:bigVis}
\end{figure}

Though most of our simulations were done for tube diameters and particles sizes that matched our experimental system, preliminary results for $\alpha=10$ in a tube with diameter twice as large show qualitatively similar behavior. Figure~\ref{fig:bigVis} shows the positions and orientations of compressive and principal stresses of each particle, and pattern appears similar to that seen in the smaller system (Fig.~\ref{fig:stressFieldVis}). It would be interesting to study the statistical distribution of tensile stresses in big homogeneous packings, to reduce possible statistical bias brought by the localized shear zone due to the applied shear deformation. It would also be interesting to characterize more precisely the transition from yielding to stiffening behavior as a function of $\alpha$.

Identifying particle shapes that exhibit enhanced or novel granular material properties suitable for practical applications is a challenging task. Our understanding of the mechanism for supporting tensile stress in the non-cohesive granular material studied here may help guide the development of composite materials with novel functionalities. For example, the formation of free-standing structures made of nonconvex particles has been considered as an alternative approach to making reinforced construction materials~\cite{damme2018_concrete}. In traditional reinforced concrete, the tensile strength is enhanced by a lattice of reinforcing steel bars (rebar). We find that the tensile stress supported by the rigid cluster in our setup is coupled to the compressive stress it receives. Moreover, the tensile yield stress of the system increases with compressive strain in an orthogonal direction, which suggests that appropriately applied compressive stresses may be used to tune the tensile strength, a feature that may prove useful for reconfigurable architectural applications.

\begin{figure}[tb]
 \centering
 \includegraphics[width=3.375 in,keepaspectratio=true]{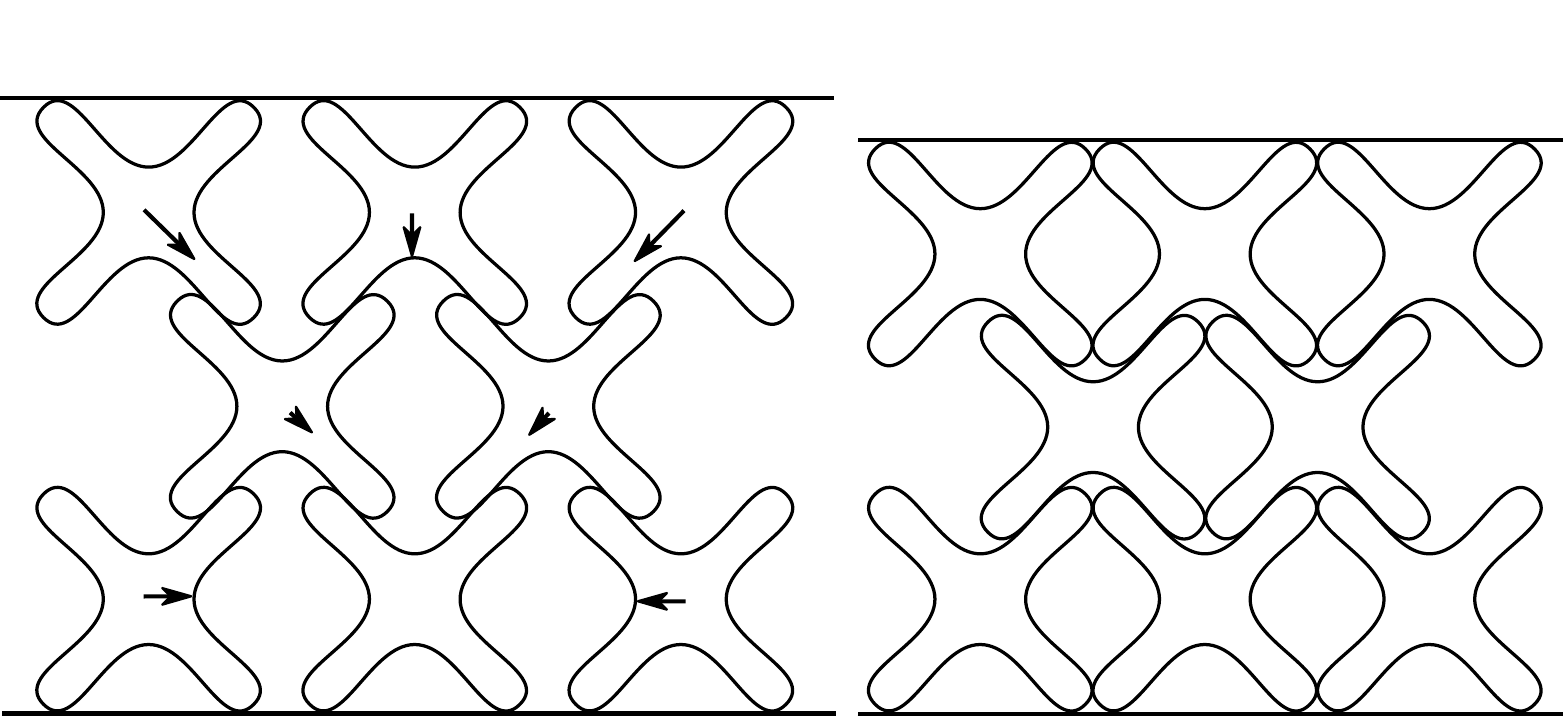}
 \caption{A structure made of crosses that contracts in the horizontal direction when compressed vertically. The structure before and after compression is shown on the left and right, respectively. Arrows indicate the displacement of individual crosses after compression.}
 \label{fig:negaP}
\end{figure}

Another possible application leveraging our insight may be a new approach to designing auxetic materials~\cite{reid2019_sm} without permanent bonds between building blocks. Figure~\ref{fig:negaP}(a) shows a lattice of rigid crosses in which contacting arms are free to slide past each other. When compressive forces are applied in the vertical direction, all contact forces generate tensile stress in the horizontal direction (Fig.~\ref{fig:tensileSimu}), causing the structure to contract horizontally as shown in Fig.\ref{fig:negaP}(b).  Different lattices and particle shapes may be used to create isotropic or anisotropic auxetic responses upon compression or extension, both in two and three dimensions.

\section*{\label{sec:acknow}ACKNOWLEDGEMENTS}
We thank Yiqiu Zhao, Ryan Kozlowski, Karen Daniels, Kieran Murphy, Heinrich Jaeger and Yo\"{e}l Forterre for helpful discussions. We also sincerely appreciate Bob Behringer for his ideas and guidance before his untimely death on July 10, 2018. This work was supported by the National Science Foundation through Grant No. DMR-1809762 and by the W. M. Keck Foundation.

\bibliography{generalbiblio}

\end{document}